\documentclass[twocolumn]{aastex63}

\usepackage{color}
\usepackage{graphicx}
\usepackage{amsmath}
\usepackage{amssymb}
\usepackage{stix}

\def\ra#1#2#3{#1$^{\rm h}$#2$^{\rm m}$#3$^{\rm s}$}
\def\dec#1#2#3{$#1^\circ#2'#3''$}

\graphicspath{{./}{figures/}}

\received{X X, XXXX}
\revised{X X, XXXX}
\accepted{July, 2020}

\submitjournal{ApJL}

\shorttitle{DRAFT}
\shortauthors{Paterson et al.}

\begin{document}
\sloppy

\title{Discovery of the optical afterglow and host galaxy of short GRB\,181123B at $z =1.754$: Implications for Delay Time Distributions}

\correspondingauthor{K. Paterson}
\email{kerry.paterson@northwestern.edu}

\newcommand{\NU}{\affiliation{Center for Interdisciplinary Exploration and Research in Astrophysics (CIERA) and Department of Physics and Astronomy, Northwestern University, Evanston, IL 60208, USA}}
\newcommand{\CfA}{\affiliation{Center for Astrophysics\:$|$\:Harvard \& Smithsonian, 60 Garden St. Cambridge, MA 02138, USA}}
\newcommand{\OU}{\affiliation{Astrophysical Institute, Department of Physics and Astronomy, 251B Clippinger Lab, Ohio University, Athens, OH 45701, USA}}
\newcommand{\Adler}{\affiliation{The Adler Planetarium, Chicago, IL 60605, USA}}
\newcommand{\GeminiN}{\affiliation{Gemini Observatory/NSF's NOIRLab, 670 N. A'ohoku Place, Hilo, HI, 96720, USA}}
\newcommand{\GSFC}{\affiliation{Astroparticle Physics Laboratory, NASA Goddard Space Flight Center, Mail Code 661, Greenbelt, MD 20771, USA}}
\newcommand{\UMD}{\affiliation{Joint Space-Science Institute, University of Maryland, College Park, MD 20742, USA}}
\newcommand{\GWU}{\affiliation{Department of Physics, The George Washington University, Washington, DC 20052, USA}}
\newcommand{\Leicester}{\affiliation{School of Physics and Astronomy, University of Leicester, University Road, Leicester, LE1 7RH, UK}}
\newcommand{\Marin}{\affiliation{College of Marin, 120 Kent Avenue, Kentfield 94904 CA, USA}}
\newcommand{\UVI}{\affiliation{University of the Virgin Islands, \#2 Brewers bay road, Charlotte Amalie, 00802 USVI, USA}}
\newcommand{\Radboud}{\affiliation{Department of Astrophysics/IMAPP, Radboud University, 6525 AJ Nijmegen, The Netherlands}}
\newcommand{\Warwick}{\affiliation{Department of Physics, University of Warwick, Coventry, CV4 7AL, UK}}
\newcommand{\Birmingham}{\affiliation{Birmingham Institute for Gravitational Wave Astronomy and School of Physics and Astronomy, University of Birmingham, Birmingham B15 2TT, UK}}
\newcommand{\Edinburgh}{\affiliation{Institute for Astronomy, University of Edinburgh, Royal Observatory, Blackford Hill, EH9 3HJ, UK}}
\newcommand{\Caltech}{\affiliation{Cahill Center for Astrophysics, California Institute of Technology, 1200 E. California Blvd. Pasadena, CA 91125, USA}}
\newcommand{\Bath}{\affiliation{Department of Physics, University of Bath, Claverton Down, Bath, BA2 7AY, UK}}
\newcommand{\Einstein}{\altaffiliation{NASA Einstein Fellow}}
\newcommand{\LJMU}{\affiliation{Astrophysics Research Institute, Liverpool John Moores University, 146 Brownlow Hill, Liverpool L3 5RF, UK}}
\author[0000-0001-8340-3486]{K. Paterson}
\NU

\author[0000-0002-7374-935X]{W. Fong}
\NU

\author[0000-0002-2028-9329]{A. Nugent}
\NU

\author[0000-0003-3937-0618]{A. Rouco Escorial}
\NU

\author[0000-0001-6755-1315]{J. Leja}
\CfA

\author[0000-0003-1792-2338]{T. Laskar}
\Bath

\author[0000-0002-7706-5668]{R. Chornock} 
\OU

\author[0000-0001-9515-478X]{A.~A. Miller}
\NU\Adler

\author[0000-0003-1585-9486]{J. Scharw\"{a}chter}
\GeminiN

\author[0000-0003-1673-970X]{S.~B.~Cenko} 
\GSFC\UMD

\author[0000-0001-8472-1996]{D.~Perley} 
\LJMU

\author[0000-0003-3274-6336]{N. R. Tanvir} 
\Leicester

\author[0000-0001-7821-9369]{A. Levan} 
\Radboud\Warwick

\author{A. Cucchiara} 
\Marin\UVI

\author[0000-0002-9118-9448]{B. E. Cobb}
\GWU

\author[0000-0002-8989-0542]{K. De}
\Caltech

\author{E. Berger} 
\CfA

\author{G. Terreran}
\NU

\author[0000-0002-8297-2473]{K. D. Alexander}
\Einstein\NU

\author[0000-0002-2555-3192]{M. Nicholl} 
\Birmingham\Edinburgh

\author{P. K. Blanchard}
\NU

\author{D. Cornish}
\NU

\begin{abstract}
We present the discovery of the optical afterglow and host galaxy of the {\it Swift} short-duration gamma-ray burst, GRB\,181123B. Observations with Gemini-North starting at $\approx 9.1$~hr after the burst reveal a faint optical afterglow with $i\approx25.1$~mag, at an angular offset of 0.59 $\pm$ 0.16$''$ from its host galaxy. Using $grizYJHK$ observations, we measure a photometric redshift of the host galaxy of $z = 1.77^{+0.30}_{-0.17}$. From a combination of Gemini and Keck spectroscopy of the host galaxy spanning 4500-18000~\AA , we detect a single emission line at 13390~\AA, inferred as H$\beta$ at $z = 1.754 \pm 0.001$ and corroborating the photometric redshift. The host galaxy properties of GRB\,181123B are typical to those of other SGRB hosts, with an inferred stellar mass of $\approx 1.7 \times 10^{10}\,M_{\odot}$, mass-weighted age of $\approx 0.9$~Gyr and optical luminosity of $\approx 0.9L^{*}$. At $z=1.754$, GRB\,181123B is the most distant secure SGRB with an optical afterglow detection, and one of only three at $z>1.5$. Motivated by a growing number of high-$z$ SGRBs, we explore the effects of a missing $z>1.5$ SGRB population among the current {\it Swift} sample on delay time distribution models. We find that log-normal models with mean delay times of $\approx 4-6$~Gyr are consistent with the observed distribution, but can be ruled out to $95\%$ confidence with an additional $\approx1-5$~{\it Swift} SGRBs recovered at $z>1.5$. In contrast, power-law models with $\propto$ $t^{-1}$ are consistent with the redshift distribution and can accommodate up to $\approx30$ SGRBs at these redshifts. Under this model, we predict that $\approx 1/3$ of the current {\it Swift} population of SGRBs is at $z>1$. The future discovery or recovery of existing high-$z$ SGRBs will provide significant discriminating power on their delay time distributions, and thus their formation channels.
\end{abstract}


\section{Introduction} \label{sec:intro}
Short-duration (T$_{90} <$ 2s) gamma-ray bursts (SGRBs) have long been linked to binary neutron star (BNS) and possibly neutron star black hole (NSBH) mergers through indirect observational evidence: the lack of associated supernovae (SNe; \citealt{Fox2005,Hjorth2005a,Hjorth2005b,Kocevski2010,Berger2014}), host galaxy demographics demonstrating a mix of young and old stellar populations \citep{Berger2009,LeiblerBerger2010,Fong2013}, low inferred environmental densities \citep{Soderberg2006,Fong2015}, moderate to large offsets with respect to their host galaxies \citep{Fong2010,Church2011,FongBerger2013, Berger2010, Tunnicliffe2014}, and emission consistent with expectations for $r$-process kilonovae \citep{Berger2013,Tanvir2013,Tanaka2016,Metzger2017,Gompertz2018,Hotokezaka2018,Lamb2019,Troja2019}. The discovery of the BNS merger GW170817 \citep{Abbott2017c} and the associated SGRB\,170817A \citep{Abbott2017a,Goldstein2017,Savchenko2017} provided direct evidence that at least some SGRBs originate from BNS mergers.

As gravitational wave (GW) facilities continue to make ground-breaking discoveries of the first BNS mergers to $z\approx 0.05$ \citep{Abbott2017c,Abbott2020}, SGRBs provide cosmological analogues that can probe the binary merger progenitor population, their rates and evolution to $z \approx 2$. Since 2004, the Neil Gehrels {\it Swift} Observatory \citep{Gehrels2004} has discovered $>130$~SGRBs \citep{Lien2016}. Dedicated campaigns to characterize their host galaxies have led to secure redshift determinations for $\approx 1/3$ of the population, with a peak in the distribution at $z\approx 0.5$ \citep{Fong2013,Berger2014,Fong2017}. However, to date, only $\approx 5\%$ of bursts have confirmed redshifts of $z>1$. This drops to $\approx 1.5\%$ (2~events) when considering confirmed secure SGRBs at $z>1.5$, GRB\,111117A at $z=2.211$ \citep{Selsing2018} and GRB\,160410A at $z=1.717$ \citep{GCN19274}\footnote{In the sample of secure SGRBs, we include events with $T_{90}<2$~s, but exclude cases like GRB\,090426A, a short-duration GRB with $\gamma$-ray and environmental properties that otherwise are more consistent with a massive star progenitor \citep{Antonelli2009,Levesque2010}.}. In general, high-redshift SGRBs are particularly challenging to characterize for a number of reasons. First, a redshift typically requires detection of an optical afterglow for sub-arcsecond precision localization and association to a host galaxy, and typical afterglow luminosities scaled to $z>1$ have faint apparent magnitudes of $r>24$~mag within hours of burst detection. Second, the sensitivity of {\it Swift} is known to fall off at higher redshifts for SGRBs compared to LGRBs due to the different thresholds in the respective detection channels \citep{GuettaPiran2005}. Third, there are a number of SGRBs with host galaxies that have inferred redshifts of $z>1.2$ due to their featureless optical spectra (e.g., GRB\,051210; \citealt{Berger2007}), but are too faint to characterize further with current near-infrared (NIR) capabilities. Finally, a broad distribution of delay times (the timescale including the stellar evolutionary and merger timescales) spanning one to several Gyr results in an event rate that will peak at low redshifts.

Given the observational challenges, the discovery of additional, confirmed SGRBs at $z>1.5$ may provide significant constraining power on the delay time distribution (DTD). In turn, the DTD inferred from SGRBs can be directly linked to the formation channel of BNS mergers, as primordial binaries versus dynamical assembly within globular clusters will result in different DTDs \citep{Hopman2006,Belczynski2018}. In the absence of other mechanisms, the merger timescale is determined by the loss of energy and angular momentum due to GW \citep{Peters1964} which can be tied to the parameters of the binary (e.g., initial separation, ellipticity; \citealt{Postnov2014,Selsing2018}). Many studies have constrained the SGRB DTD by fitting the SGRB redshift distribution, predominantly focused on the $z<1$ population \citep{Nakar2006,Berger2007,Jeong2010,Hao2013,Wanderman2015,Anand2018}. Other constraints on the DTD have come from studies on the Galactic population of NS binaries \citep{Champion2004,Beniamini2016,VignaGomez2018,Beniamini2019}, and SGRB host galaxy demography, as longer delay times will result in an increase in host galaxies with old stellar populations \citep{Zheng2007,Fong2013,Behroozi2014}.

Here, we present the discovery of the optical afterglow and host galaxy of GRB\,181123B at $z=1.754$, making this event the third confirmed event at $z>1.5$, and the most distant, secure SGRB with an optical afterglow to date. In Section \ref{sec:GRB181123B} we describe the discovery and community observations of GRB\,181123B. We describe our photometric and spectroscopic observations of GRB\,181123B in Section \ref{sec:obs}. In Sections \ref{sec:burst} and \ref{sec:host}, we discuss the burst explosion and host galaxy properties, respectively. Our results, including a discussion of GRB\,181123B in context with the population of SGRBs and the implications on the DTD, is given in Section \ref{sec:discussion}. Finally we summarize our conclusions in Section \ref{sec:conclusion}.

Unless otherwise stated, all observations are reported in AB mag and have been corrected for Galactic extinction in the direction of the burst \citep{Schlafly2011}. We employ a standard cosmology of $H_{0}$ = 69.6, $\Omega_{M}$ = 0.286, $\Omega_{vac}$ = 0.714 \citep{Bennett2014}.

\section{GRB 181123B} \label{sec:GRB181123B}
On 2018 Nov 23 at 05:33:03 UT the Burst Alert Telescope (BAT; \citealt{Barthelmy2005}) on board the Neil Gehrels \textit{Swift} Observatory \citep{Gehrels2004} discovered and located GRB\,181123B at a refined position RA, Dec = \ra{12}{17}{27.99}, \dec{14}{35}{56.0} (3.8$''$ positional uncertainty; 90\% confidence) with a duration of $T_{90}$ = 0.26 $\pm$ 0.04s in the 50-300 keV band (90\% confidence; \citealt{gcn23432}). GRB\,181123B also showed minimal spectral lag \citep{gcn23443} and a hardness ratio (between the 50-100 and 25-50 keV bands) of 2.4 \citep{Lien2016}. These properties classify GRB\,181123B as a short-duration, hard spectrum GRB. A power-law fit to the data results in a fluence, f$_{\gamma}$ = ($1.2 \pm 0.2$) $\times 10^{-7}$ erg cm$^{-2}$ in the 15-150 keV band \citep{Evans2009}. \textit{Swift}'s X-ray Telescope (XRT) began observing the field at $\delta t$ = 80.25s, where $\delta t$ is the time after the BAT trigger in the observer frame, localizing an uncatalogued X-ray source within the BAT region with an enhanced position of RA, Dec = \ra{12}{17}{28.05}, \dec{14}{35}{52.4} with a positional uncertainty of 1.6$''$ (90\% confidence; \citealt{gcn23434}, \citealt{Goad2007}, \citealt{Evans2009}). Follow-up observations performed by {\it Swift}'s Ultraviolet/Optical Telescope (UVOT) resulted in no afterglow detection to a 3$\sigma$ upper limit of V $>$ 19.7 mag at mid-time of $\delta t$ = 5148 s \citep{gcn23437}.

From the community, Mobile Astronomical System of Telescope-Robots (MASTER, \citealt{Lipunov2010}) obtained optical follow-up observations at $\delta t$ = 2.7 and 20.2 hrs and did not detect any source in or around the XRT position to upper limits of $\gtrsim$ 17 and $\gtrsim$ 18.1 mag respectively \citep{gcn23444}. In addition, radio observations with the Australia Telescope Compact Array (ATCA, \citealt{Frater1992}) were conducted at $\delta t$ = 12.5 hrs; no radio emission was detected to 3$\sigma$ upper limits of 66 and 69 $\mu$Jy at 5 and 9 GHz respectively \citep{gcn23467}.

\begin{deluxetable*}{cccccccccc}[t!]
\tablecaption{Afterglow and Host Galaxy Observations of GRB\,181123B \label{tab:obs_log}}
\tablecolumns{7}
\tablewidth{0pt}
\tablehead{
\colhead{Date$^\dagger$} &
\colhead{$\delta t^\dagger$} &
\colhead{Filter} & \colhead{Telescope} & \colhead{Instrument} &
\colhead{Total exp. time} & \colhead{Afterglow} & \colhead{Host} & \colhead{A$_{\lambda}^*$}\\
\colhead{(UT)} & \colhead{(days)} & \colhead{} &
\colhead{} & \colhead{} & \colhead{(s)} & \colhead{(AB mag)} & \colhead{(AB mag)} & \colhead{(AB mag)}
}
\startdata
 & & & & \textit{Imaging} & & & & \\
2018 Nov 23.618 & 0.38 & $i$ & Gemini-N & GMOS & 1620 & 25.10 $\pm$ 0.39 & \nodata & 0.06\\
2018 Nov 23.669 & 0.43 & $J$ & Keck I & MOSFIRE & 2880.5 & $>$ 23.2 & 23.05 $\pm$ 0.32 & 0.03\\
2018 Nov 25.637 & 2.40 & $i$ & Gemini-N & GMOS & 2400 & \nodata & 23.79 $\pm$ 0.19 & 0.06\\
2018 Nov 25.496 & 2.27 & $J$ & MMT & MMIRS & 3717.7 & $>$ 23.3 & 22.65 $\pm$ 0.26 & 0.03\\
2018 Nov 26.655 & 3.42 & $J$ & Keck I & MOSFIRE & 2138.5 & \nodata & 22.85 $\pm$ 0.23 & 0.03\\
2019 Jan 17.545 & 55.31 & $K$ & MMT & MMIRS & 1704 & \nodata & 22.33 $\pm$ 0.23 & 0.01\\
2019 Jan 20.464 & 58.23 & $K$ & MMT & MMIRS & 1394.2 & \nodata & 22.39 $\pm$ 0.41 & 0.01\\
2019 Feb 03.312 & 72.08 & $r$ & Gemini-S & GMOS & 1800 & \nodata & 23.84 $\pm$ 0.19 & 0.08\\
2019 Feb 03.340 & 72.11 & $g$ & Gemini-S & GMOS & 1800 & \nodata & 24.08 $\pm$ 0.23 & 0.12\\
2019 Feb 03.370 & 72.14 & $z$ & Gemini-S & GMOS & 1800 & \nodata & 23.84 $\pm$ 0.22 & 0.04\\
2019 May 24.275 & 182.04 & $H$ & MMT & MMIRS & 2987.4 & \nodata & 22.61 $\pm$ 0.19 & 0.02\\
2020 March 5.329 & 468.10 & $Y$ & MMT & MMIRS & 3584.9 & \nodata & 22.78 $\pm$ 0.24 & 0.03\\
\hline
 & & & & \textit{Spectroscopy} & & & & \\
2019 Feb 26.495 & 95.26 & $GG455$ & Keck II & DEIMOS & 5400 & & & \\
2019 Apr 10.678 & 137.95 & $JH$ & Gemini-S & FLAMINGOS-2 & 3600  & & & \\
\enddata
\tablecomments{All magnitudes are in the AB system and uncertainties correspond to $1\sigma$. \\
$^\dagger$ Based on mid-time of observation. \\
$^*$ Galactic extinction in the direction of the burst \citep{Schlafly2011}.}
\end{deluxetable*}

\section{Observations} \label{sec:obs}
\subsection{Afterglow observations} \label{sec:afterglow_phot}
\subsubsection{Gemini optical discovery}
We initiated $i$-band ToO observations of the field of GRB\,181123B with the Gemini Multi-Object Spectrograph (GMOS, \citealt{Crampton2000}) mounted on the 8-m Gemini-North telescope (PI Fong, Program GN-2018B-Q-117), at a mid-time of 2018 Nov 23.618 UT or $\delta t = 9.2$~hr \citep{gcn23439}. We obtained $18 \times 90$~s of exposures, resulting in a total of 1620 s on source, at an airmass of $1.7$, and average seeing of $1.0''$. We used a custom pipeline\footnote{\url{https://github.com/KerryPaterson/Imaging_pipelines/GMOS_pipeline.py}}, using routines from \texttt{ccdproc} \citep{Craig2017} and \texttt{astropy} (\citealt{astropy:2013}, \citealt{astropy:2018}) to perform bias subtraction, flat-fielding and gain correction calibrations. We aligned and co-added the data and performed astrometry relative to \textit{Gaia} DR2 (\citealt{Gaia2016}, \citealt{Gaia2018}, \citealt{Lindegren2018}).

We obtained a second, deeper set of $i$-band Gemini-N/GMOS observations at a mid-time of 2018 Nov 25.637 UT ($\delta t = 2.41$~days) with significantly improved image quality compared to the first set (airmass of $1.3$, seeing of $0.7''$). We detect an extended source in the epoch 2 observations near the XRT position, presumed to be the host galaxy (see Section~\ref{sec:host_obs}). To assess any fading between epoch 1 and epoch 2, we align the epoch 2 observations with respect to epoch 1 and perform image subtraction using HOTPANTS (\citealt{Becker2015}, Figure \ref{fig:subtraction}) between the two epochs. A source is found within the enhanced XRT position in the difference image, which we consider to be the optical afterglow. Performing aperture photometry with standard \texttt{IRAF} \citep{Tody1986} packages directly on the residual image and photometrically calibrating to the Sloan Digital Sky Survey (SDSS; \citealt{Fukugita1996}, \citealt{Ahn2012}, \citealt{Eisenstein2011}, \citealt{Gunn2006}), we measure an afterglow brightness of $i$ = 25.10 $\pm$ 0.39 mag at $\delta t$ = 9.2 hrs. Calibrated to \textit{Gaia} DR2, we determine a position at RA, Dec = \ra{12}{17}{27.94}, \dec{14}{35}{52.66} with positional uncertainty of 0.10$''$ accounting for the afterglow centroid and astrometric uncertainties. A summary of our observations and aperture photometry is given in Table \ref{tab:obs_log}.

\subsubsection{NIR photometric observations}
We obtained $J$-band observations of GRB\,181123B with the Multi-Object Spectrometer For Infra-Red Exploration (MOSFIRE, \citealt{McLean2012}) mounted on the 10-m Keck I telescope (PI Miller, Program 2018B\_NW254), at a mid-time of 2018 Nov 23.669 UT or $\delta t = 10.4$~hr (first reported in \citealt{gcn23440}). We obtained a total of 2880.5 s on source, from $36 \times 58$ s exposures and $27 \times 29$ s exposures, in cloudy conditions with $1''$ seeing. We developed and used a custom MOSFIRE pipeline\footnote{\url{https://github.com/KerryPaterson/Imaging_pipelines/MOSFIRE_pipeline.py}} (using routines from \texttt{ccdproc} and \texttt{astropy}) to reduce the data in a similar manner to Gemini, but with an additional sky subtraction routine to take into account the varying IR sky. We aligned and co-added the data, dividing first by the exposure time to ensure equal weights, and performed astrometry relative to \textit{Gaia} DR2.

\begin{figure*}
\centering
\includegraphics[width=0.8\textwidth]{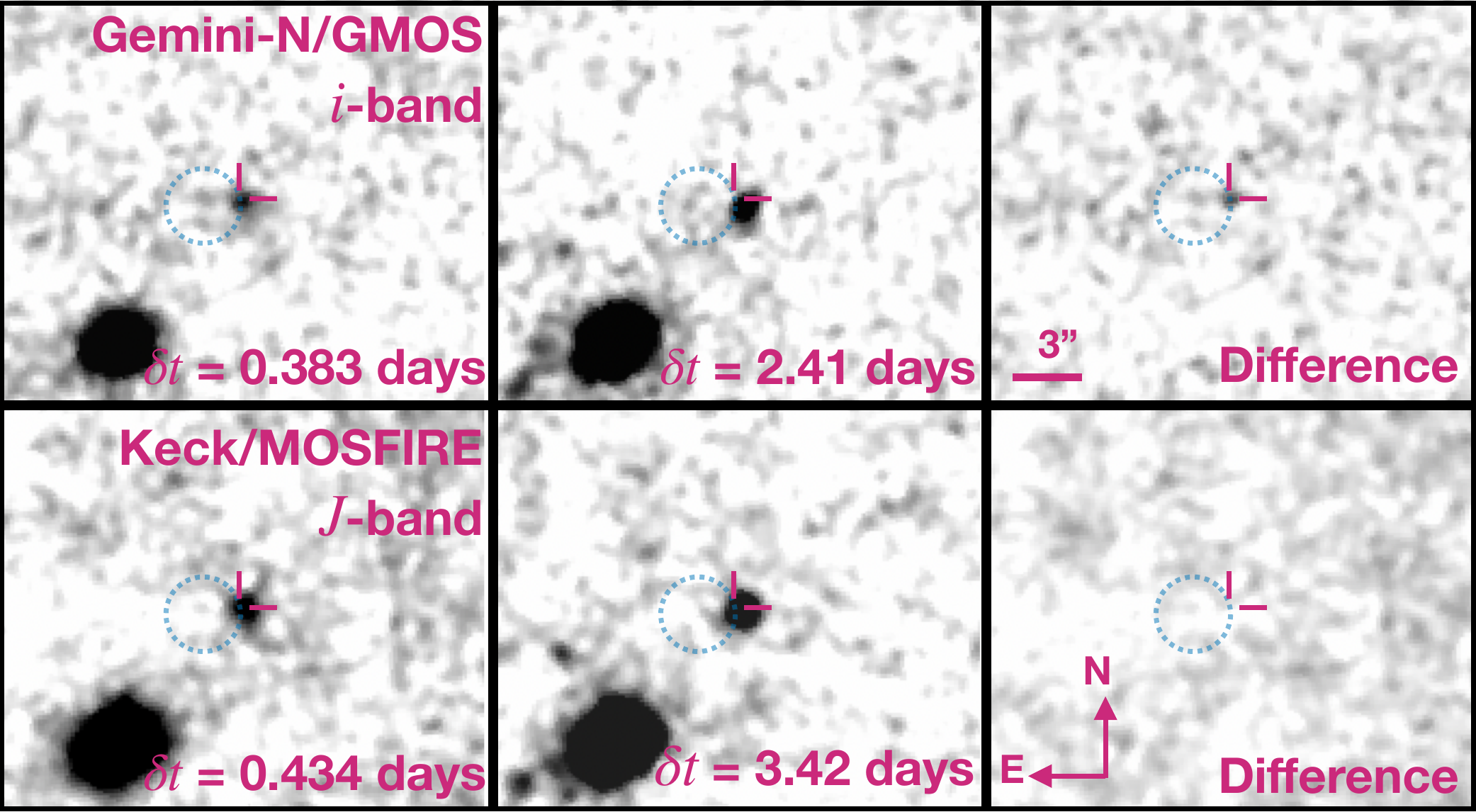}
\caption{First epoch (left) and second epoch (middle) of $i$ and $J$-band observations, and the difference between the two images produced using HOTPANTS (right), smoothed for display purposes. The circle shows the 90\% XRT localization of GRB\,181123B. The $i$-band difference image reveals a faint, $i=25.10 \pm 0.39$~mag source on the edge of the XRT position, whose position is marked by the crosshair, which we consider to be the optical afterglow. We find no source in the $J$-band difference image to a limit of $J \gtrsim 23.2$~mag. \label{fig:subtraction}}
\end{figure*}

We detect an extended source near the XRT position, and initiated a second set of observations with Keck \citep{gcn23461} through a ToO program (PI Fong, Program 2018B\_NW249) at a mid-time of 2018 Nov 26.655 UT or $\delta t = 3.42$ days. We obtained $40 \times 58$ s exposures, for a total of 2138.5 s on source, in clear conditions with seeing of $0.9''$. The clear conditions and improved seeing of these observations provided a deeper image, allowing us to use it as a template for image subtraction. We align this image with the epoch 1 observations and perform image subtraction using HOTPANTS. We do not detect any residuals at the position of the afterglow to a 3$\sigma$ limit of $J\gtrsim23.2$~mag, calibrated to 2MASS \citep{Skrutskie2006} and converted to the AB system. The images are shown in Figure \ref{fig:subtraction}.

We also obtained $J$-band observations with the Magellan Infrared Spectrograph (MMIRS, \citealt{McLeod2012}) mounted on the 6.5-m MMT telescope (PI Fong, Program 2018C-UAO-G4) at $\delta t$ = 2.27 days. We developed and used a custom MMIRS pipeline\footnote{\url{https://github.com/KerryPaterson/Imaging_pipelines/MMIRS_pipeline.py}} to reduce data in a similar manner to MOSFIRE. We perform image subtraction relative to the second epoch of Keck, and do not detect any residuals at the position of the afterglow to a 3$\sigma$ limit of $J \gtrsim 23.3$~mag, calibrated to 2MASS and converted to the AB system. We note that the lack of afterglow detection in $J$-band is consistent with the steady brightness of the host galaxy over all three epochs (Table~\ref{tab:obs_log}).

\subsection{Host observations}
\label{sec:host_obs}

\subsubsection{Host Galaxy Assignment}
We quantify the probability that the coincident extended source is the host galaxy of GRB\,181123B. Based on the XRT position alone, we calculate the probability of chance coincidence (P$_{\rm{cc}}$, \citealt{Bloom2002}) of the GRB with the galaxy to be $P_{\rm cc}=0.012$. From Gemini $i$-band imaging, we measure an afterglow brightness of $i$ = 23.85 $\pm$ 0.19 mag and determine a position of RA, Dec = \ra{12}{17}{27.91}, \dec{14}{35}{52.27} with a positional uncertainty of 0.07$''$. Relative to the optical afterglow position, we calculate an offset of 0.59 $\pm$ 0.16$''$, taking into account the afterglow and host centroids and relative astrometric uncertainty. Using the optical afterglow, we calculate P$_{\rm{cc}} = 4.4 \times 10^{-3}$. Calculating a P$_{\rm{cc}}$ for nearby extended sources in the field, the next most probable host has P$_{\rm{cc}}$ = 0.07, while all other sources have values close to unity. Thus, we conclude that the extended source is the host galaxy of GRB\,181123B.

\subsubsection{Multi-band Imaging} \label{sec:host_phot}
We obtained late-time $g$, $r$ and $z$-band observations with Gemini-South/GMOS at $\delta t \approx$ 72 days (PI Fong, Program GS-2018B-Q-112). We also obtained $YHK$ observations, where $Y$-band observations are calibrated to UKIRT (\citealt{Hewett2006}, \citealt{Lawrence2007}, \citealt{Hodgkin2009}) and converted to the AB system, with MMT/MMIRS (PI Fong, Programs 2019A-UAO-G7 and 2020A-UAO-G212-20A) with $\delta t >$ 50 days.

\begin{figure*}
\includegraphics[width=0.5\textwidth]{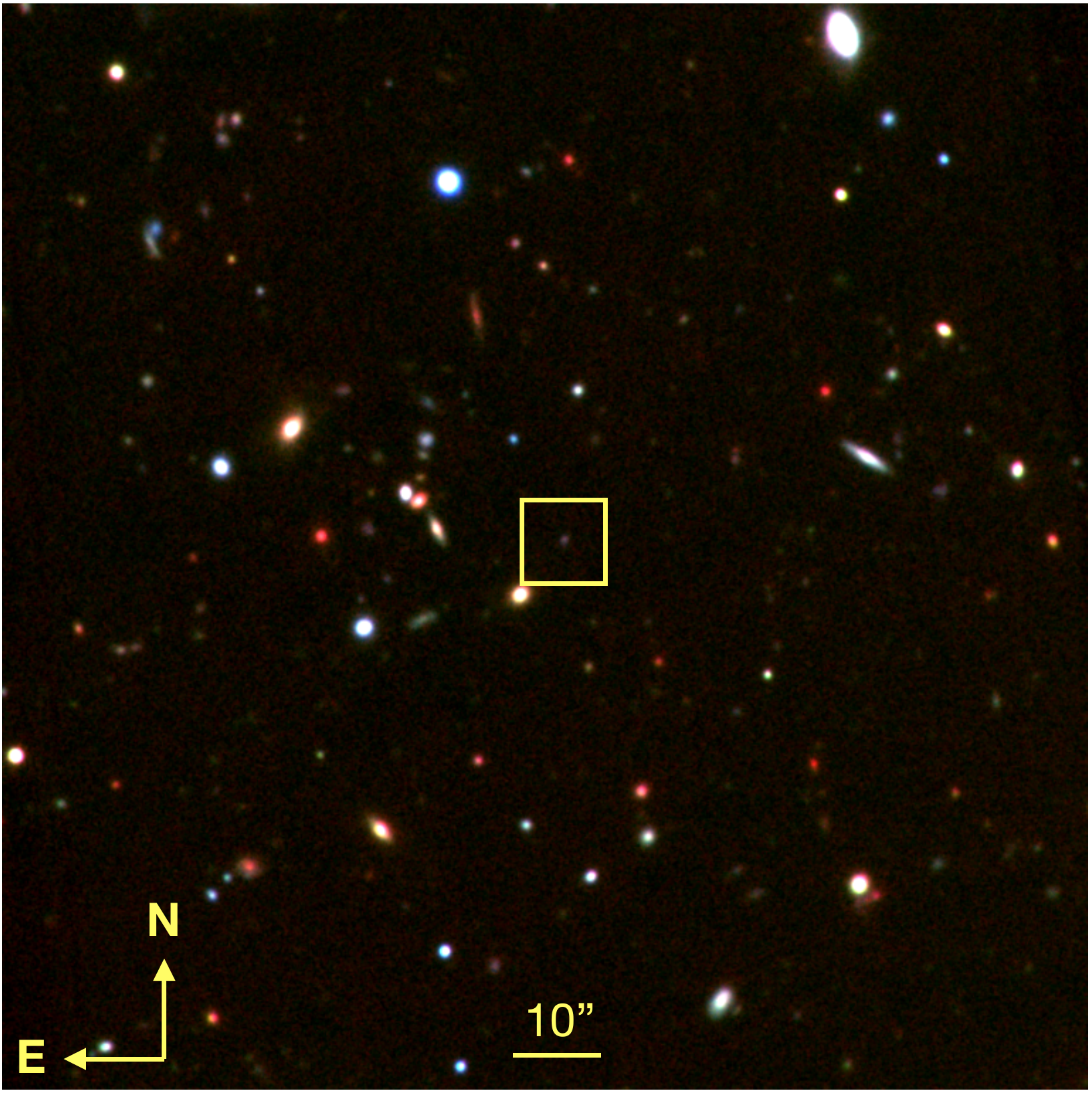}
\includegraphics[trim = 0cm -2.5cm 0cm 0cm,width=0.49\textwidth]{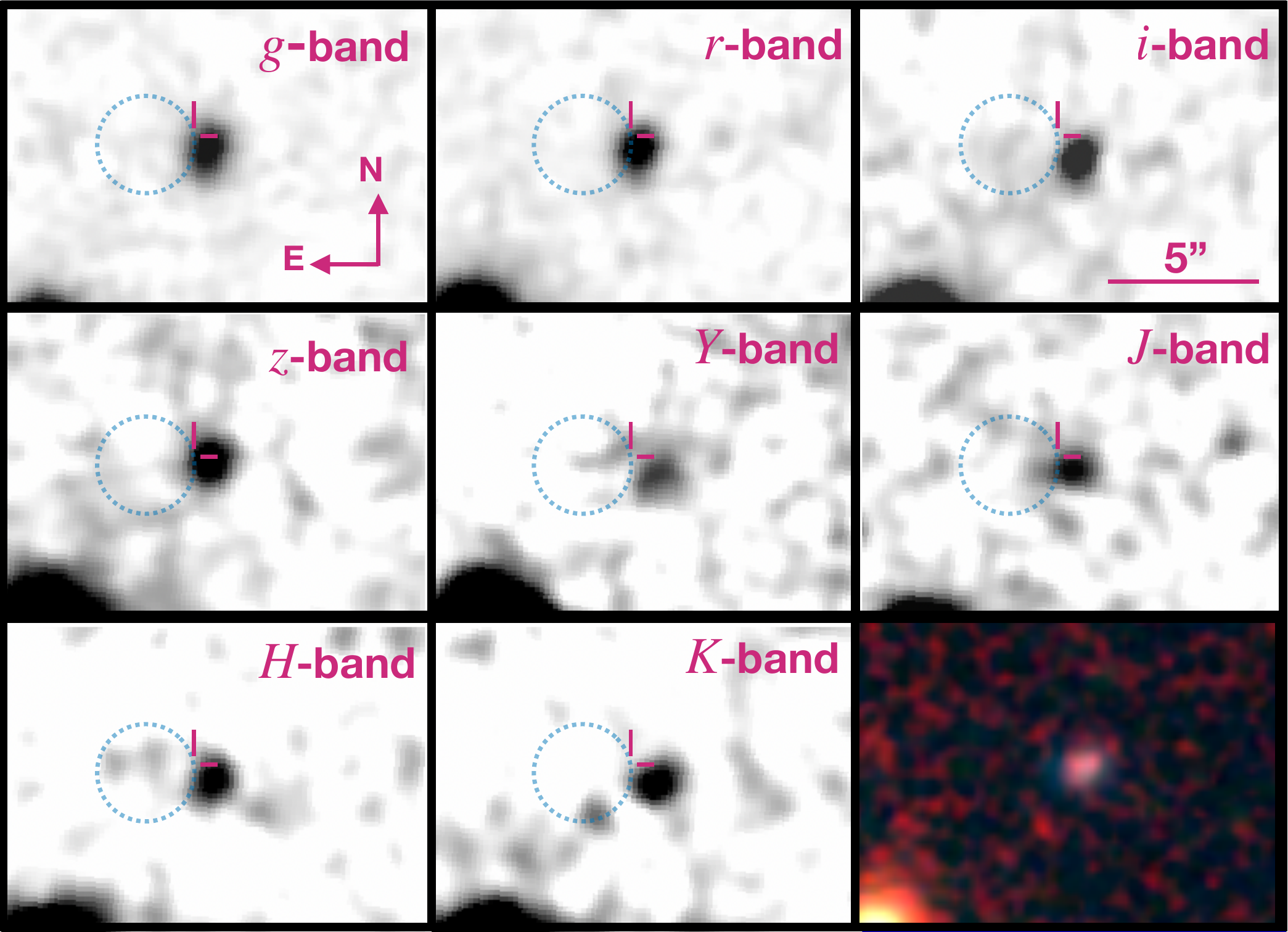}
\caption{Left: Color composite 6-filter image of GRB\,181123B observations created using \texttt{AstroImageJ} \citep{Collins2017}: $g$ = blue, $r$ = cyan, $i$ = green, $z$ = yellow, $J$ = red, $K$ = magenta showing the large field-of-view. Right: Multi-band photometry of GRB\,181123B's host galaxy (zoomed region highlighted by the yellow box of the color composite image) obtained with the MMT, Gemini and Keck; the last panel is the color composite image. The {\it Swift}/XRT position is denoted by the blue circle ($90\%$ confidence) and images are smoothed for display purposes. \label{fig:photometry}}
\end{figure*}

The details of these observations are summarized in Table \ref{tab:obs_log}. A color composite image of the field made from the $g$, $r$, $i$, $z$, $J$, and $K$ filters, along with the photometry of the host galaxy from all bands is shown in Figure \ref{fig:photometry}.

\vspace{1cm}
\subsubsection{Keck optical spectroscopy} \label{sec:host_spec}
We obtained an optical spectrum of the host of GRB\,181123B with the DEep Imaging Multi-Object Spectrograph (DEIMOS, \citealt{Faber2003}) mounted on Keck I (PI Paterson, Program 2019A\_O329). We obtained $3 \times 1800$ s exposures in clear conditions with $0.9''$ seeing. Using the 600ZD grating, a GG455 order-block filter and a central wavelength of 7498\AA, the spectrum roughly covers the wavelength range 4400 - 9600\AA\ with a central resolving power, R = 2142. We used standard \texttt{IRAF} routines in the \texttt{ctioslit} package to reduce and co-add the data. We performed wavelength calibrations using a NeArKrXe arc taken just before the observations and used the standard star Feige 34 for spectrophotometric calibration. We extracted the error spectrum and normalised by $1/\sqrt{N}$, where $N$ is the number of images used in the co-add. The resulting spectrum, scaled to the multi-band photometry is shown in Figure \ref{fig:spectrum}.

The spectrum is featureless, with no lines above a signal-to-noise ratio (SNR) $>$ 5. There is a faint continuum but no clear features. In particular, the lack of identifiable lines or features suggests that the host galaxy is at $z>1.4$.

\subsubsection{Gemini near-infrared spectroscopy}
\begin{figure*}[!t]
\includegraphics[width=0.495\textwidth]{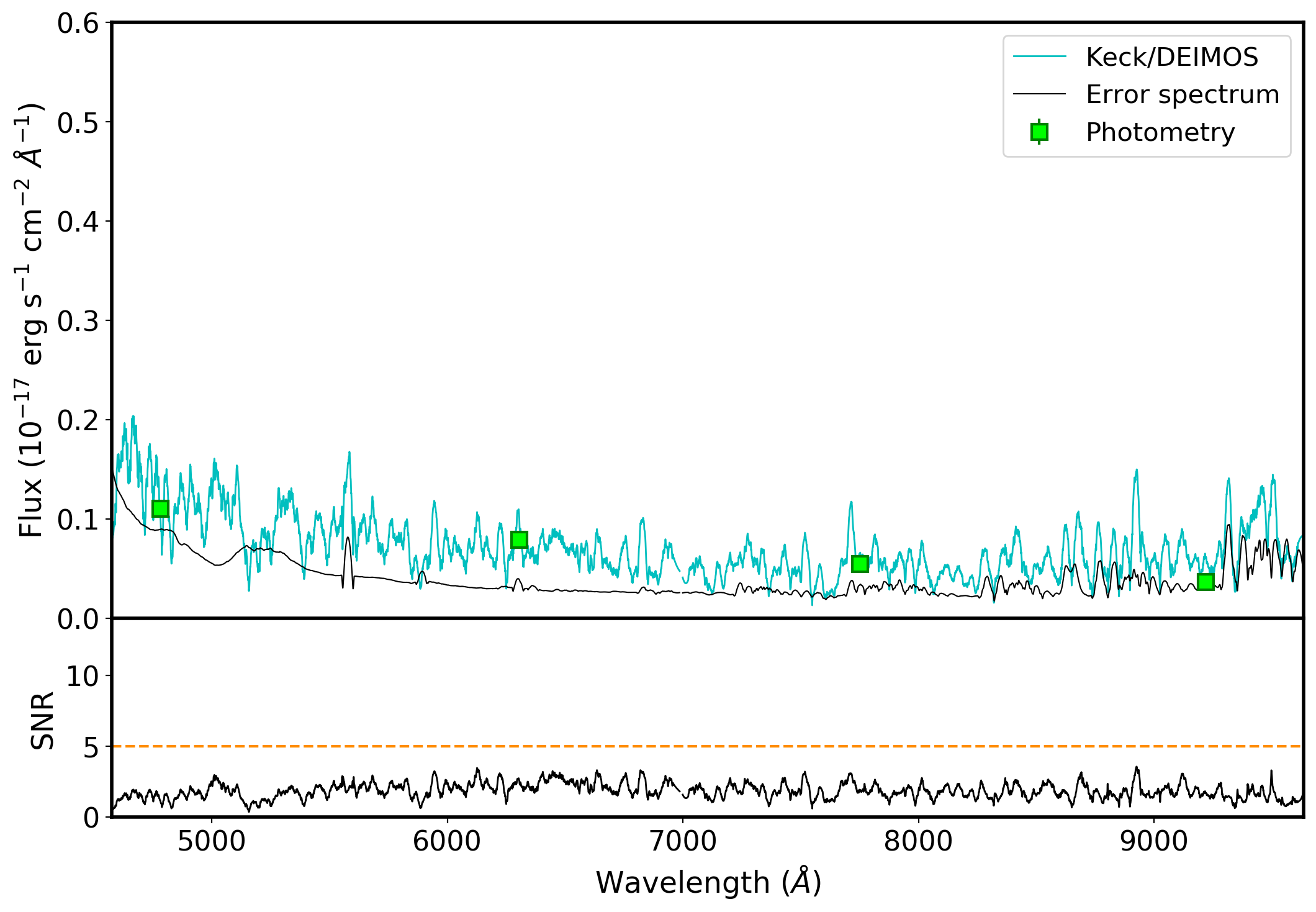}
\includegraphics[width=0.51\textwidth]{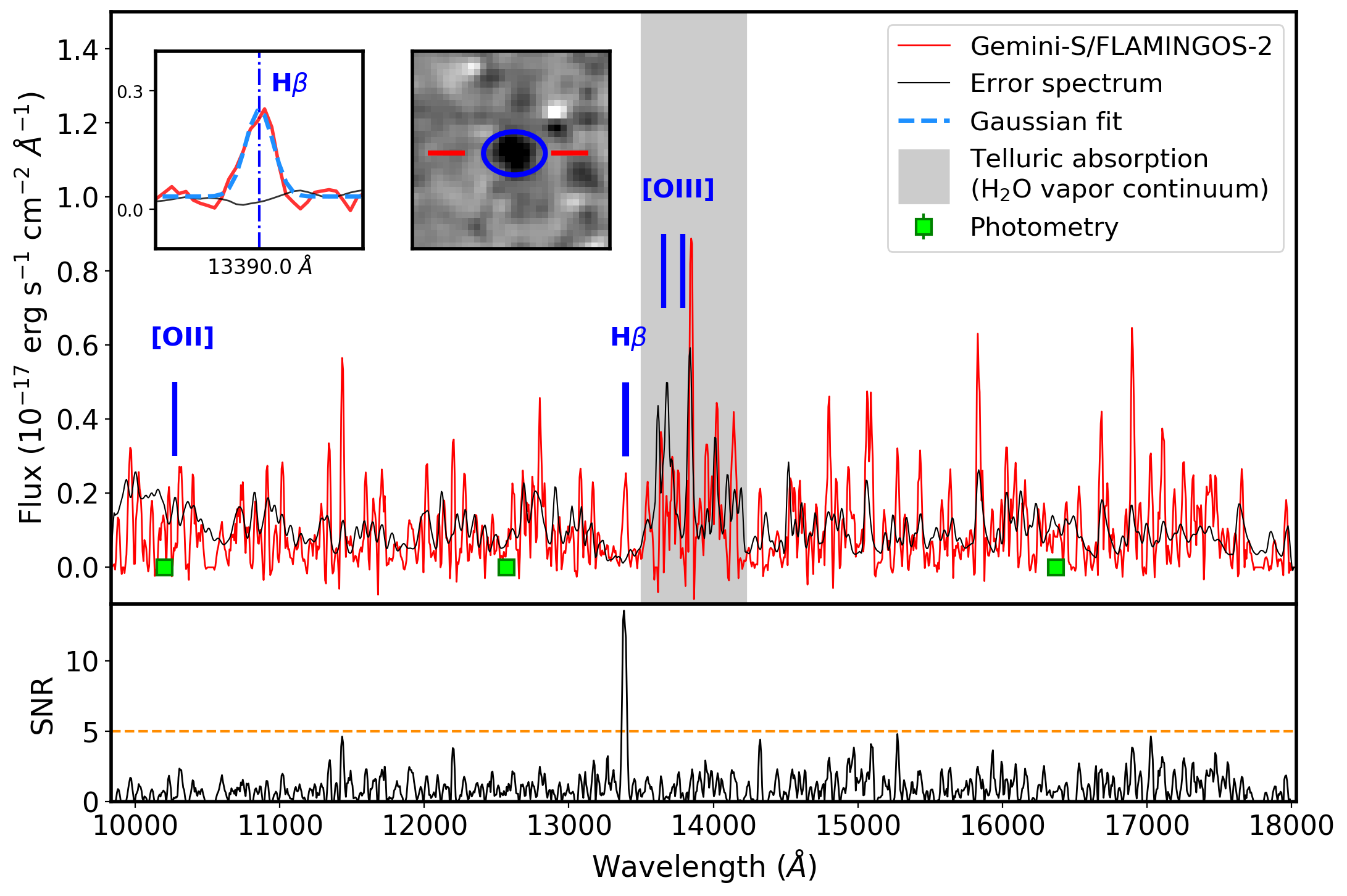}
\caption{Top: Optical (left) and NIR (right) spectrum of GRB\,181123B's host galaxy. The cyan line shows the optical spectra Keck/DEIMOS. The red line shows the NIR spectrum from Gemini-S/FLAMINGOS-2. The black lines show the error spectra. The gray band shows the region of strong telluric absorption caused by continuum water vapor absorption. The spectrum has been scaled to the photometry (green squares) and smoothed using a 75 windowed, 2nd order polynominal Savitzky-Golay filter \citep{Savitzky1964}. The spectrum and photometry are de-reddened by the Galactic extinction in the direction of the burst (\citealt{Schlafly2011}, \citealt{Cardelli1989}). The blue lines correspond to the positions of strong emission lines expected at $z$ = 1.754; the [OII] and [OIII] doublets are both in regions of large error. The inserts on the right shows a zoomed-in view on the position of the H$\beta$ line for the 1D spectrum fitted with a Gaussian (left insert) and the 2D spectrum highlighting the emission line (right insert). Bottom: SNR of the respective spectra. The orange dashed line corresponds to SNR = 5. Only a single line with SNR = 13.5 is seen in the NIR at 13390 \AA, which we identify as H$\beta$.
\label{fig:spectrum}}
\end{figure*}

We obtained NIR spectroscopy with the The Facility Near-Infrared Wide-field Imager \& Multi-Object Spectrograph for Gemini (FLAMINGOS-2, \citealt{Eikenberry2004}) mounted on the 8-m Gemini-South telescope using a fast-turnaround Program (PI Paterson, Program GS-2019A-FT-107). Using a JH/JH grism/filter setup with a central wavelength of $13900$\,\AA\, we obtained $30 \times 120$ s exposures, for a total of 3600 s on source, roughly covering the wavelength range $9800-18000$\,\AA\ and with a central R = 1177. We used standard procedures from the \texttt{gemini} package within \texttt{IRAF} to reduce and co-add the data. We performed wavelength calibration using Ar arc lamp spectra, and flux calibration and telluric line corrections with the standard star HIP56736, using the generalized IDL routine \texttt{xtellcor\_general} \citep{Vacca2003} from the \texttt{Spextool} package \citep{Cushing2004}. We extracted the error spectrum in the same manner as the Keck spectrum; the FLAMINGOS-2 spectrum, scaled to the $YJH$ photometry, is shown in Figure \ref{fig:spectrum}.

\subsubsection{Redshift determination}
We identify a single emission line in the FLAMINGOS-2 spectrum with a SNR = 13.5 at 13390.0 \AA. No other line features are present with SNR $\gtrsim$ 5. Given that this is the only clear line in the spectrum, we explore if the line can be matched to one of four possibilities based on the predominant features in star-forming (SF) galaxy spectra ([OII]$\lambda3727$, H$\beta\lambda4861$, [OIII]$\lambda4959$/$\lambda5007$ or H$\alpha\lambda6563$) given the photometric redshift of photo-$z = 1.77^{+0.30}_{-0.17}$ based on the 8-filter host photometry (discussed in Sections~\ref{sec:host_phot}) and the absence of any other features over $4400-18000$\,\AA.

If this line is H$\alpha$, the H$\beta$ and [OII] lines would fall in regions of low error and should have been detected. Similarly, if this line is either of the [OIII]$\lambda4959$/$\lambda5007$ doublet, the H$\beta$ and [OII] lines should have been detected. In this case, the resulting redshifts would be $z=1.70$ and $z=1.67$, respectively.  If the line is [OII], the resulting redshift of $z = 2.59$, which is not consistent with the photo-$z$ (see Section \ref{sec:host}) and the H$\beta$ line should have been detected. Finally, if the line is H$\beta$, the resulting redshift is fully consistent with the photo-$z$. The location of the [OIII] doublet is in a region of strong telluric absorption, the location of [OII] is in a region of high noise, and the location of H$\alpha$ is not covered. Considering that the [OII] line is not detected due to the high noise, we calculate a [OII]/H$\beta$ ratio, based on a SNR of 5 for the [OII] line and assuming a similar line width, on the order of $\lesssim$ 5. We thus determine the most likely candidate for this line is H$\beta$, which would place GRB\,181123B at $z$ = 1.754 $\pm$ 0.001. We thus use this redshift for our subsequent analysis.

\begin{figure*}[t]
\centering
\includegraphics[width=0.45\textwidth]{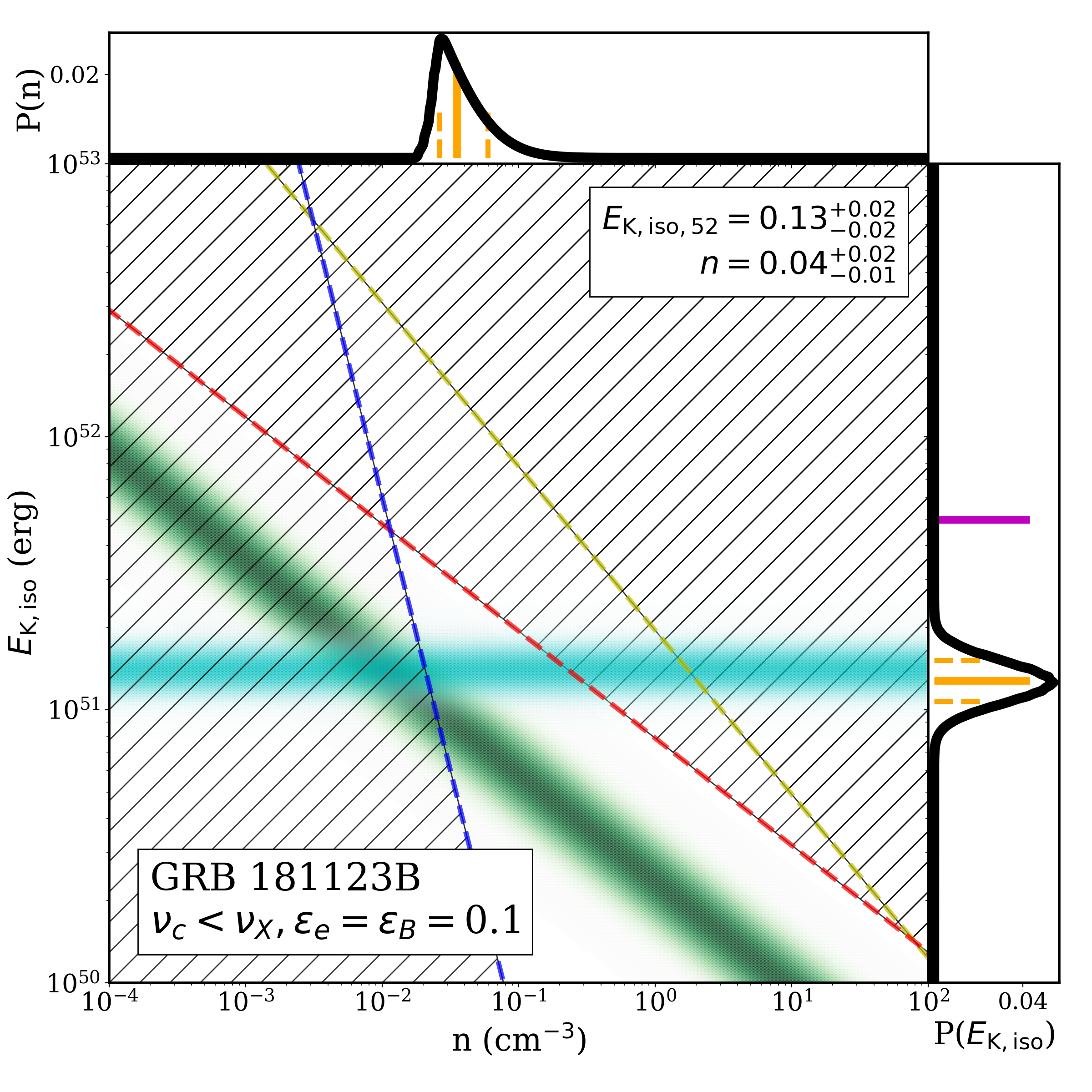}
\includegraphics[width=0.45\textwidth]{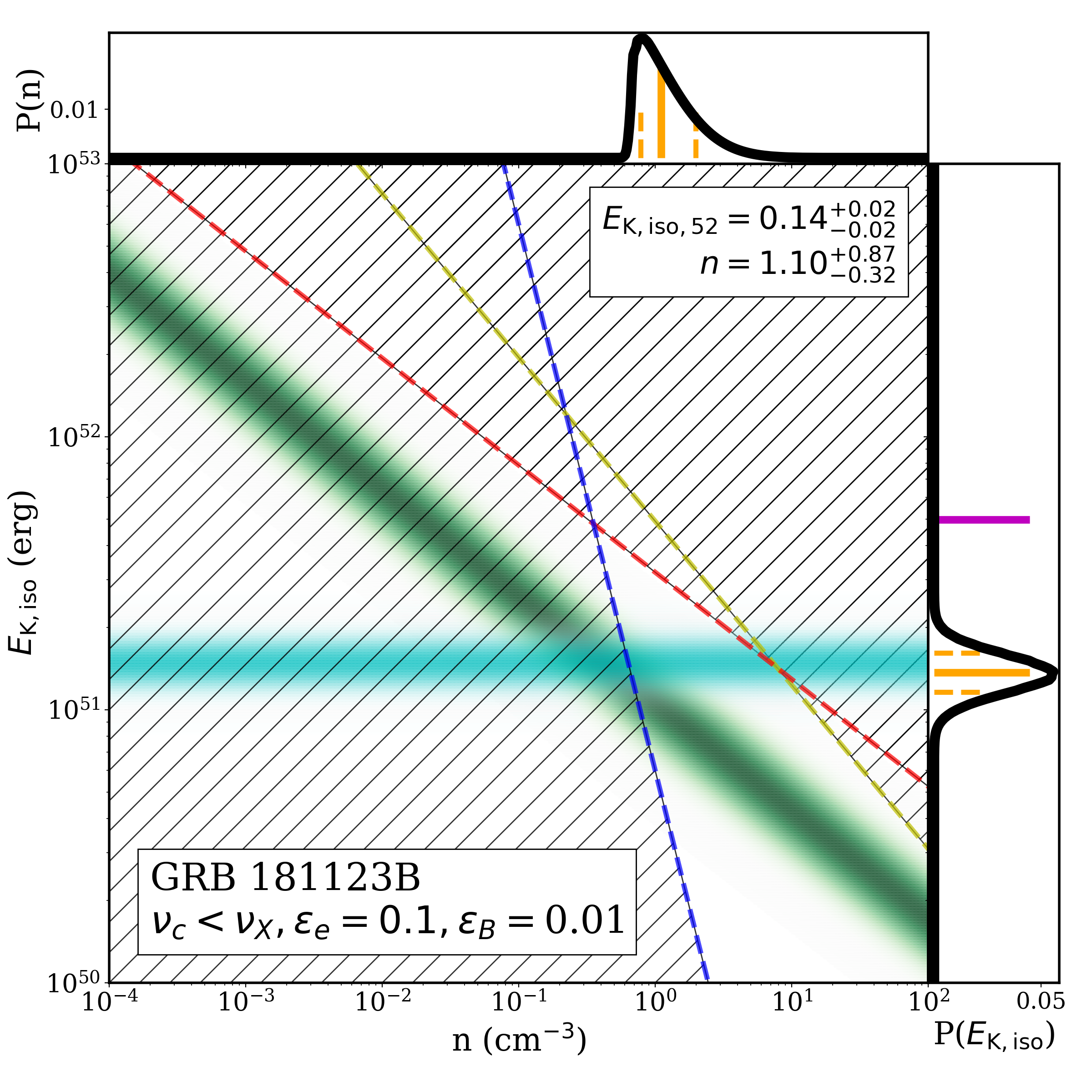}
\caption{Isotropic-equivalent kinetic energy, $E_{\rm{K,iso}}$, versus circumburst density, $n$, for the afterglow of GRB\,181123B. The solid cyan and green bands show the solution from the X-ray and optical detections, respectively, where the width corresponds to $1\sigma$ confidence. The fading in color represents the drop off in probability going away from the center of each constraint. The red, yellow, and blue lines correspond to the solutions from limits set by the NIR band, radio band, and cooling frequency, respectively, where the corresponding hatched regions illustrate the parameter space ruled out by the limits. The solid black distributions to the top and right of the parameter space shows the 1D probability of $E_{\rm{K,iso}}$ and $n$ respectively. The solid magenta line to the right shows $E_{\gamma,iso}$. Left: Case where $\epsilon_B$ = 0.1. Right: Case where $\epsilon_B$ = 0.01.
\label{fig:energy_density}}
\end{figure*}

\section{Afterglow properties} \label{sec:burst}
Adopting the standard synchrotron model for a relativistic blastwave expanding into a constant density medium (\citealt{Sari1998}, \citealt{Granot2002}), we use the broadband afterglow observations to infer physical parameters such as the electron power-law index ($p$), the isotropic-equivalent kinetic energy ($E_{\rm{K,iso}}$), circumburst density ($n$) and the fraction of electrons in the electric ($\epsilon_{e}$) and magnetic field ($\epsilon_{B}$) using the standard relations from \cite{Granot2002}. The relation of the observed flux to the physical parameters requires knowledge of the location of the spectral break frequencies with respect to the observing bands, and hence the part of the spectrum each band falls on.

For the X-rays, we download the \textit{Swift}/XRT data available from the \textit{Swift} website \citep{Evans2009}. We make use of the late-time observations for our fits due to excess flux at early times. We use the temporal and the spectral power-law indices from the X-rays ($\alpha_X$ and $\beta_X$ respectively), to determine if the X-ray band falls below or above the cooling frequency, $\nu_c$, through the calculation of $p$. Fitting a power law to the XRT light curve at $\delta t >$ 700 s using $\chi^2$-minimization, we obtain $\alpha_X = -0.96^{+0.13}_{-0.11}$. We use \texttt{XSPEC} \citep{Arnaud1996} to fit a two absorption power-law to the XRT spectrum over $\delta t = 561-16171$ s. Fixing the Galactic hydrogen column density, $N_{\rm{H,Gal}}= 3.08 \times 10^{20}$ cm$^{-2}$ \citep{Willingale2013} and $z$ = 1.754, we find an intrinsic hydrogen column density of $N_{\rm{H,int}}=8.40^{+61.10}_{-8.40} \times 10^{20}$ cm$^{-2}$ and photon index $\Gamma_X = 2.392^{+0.133}_{-0.422}$. We determine $\beta_X = -1.39^{+0.13}_{-0.42}$ from the definition $\beta_X \equiv 1 - \Gamma_X$. Using the median values for the spectral parameters, we calculate an unabsorbed X-ray flux at 2.3 hr of $F_X = 2.07 \pm 0.43$~erg~s$^{-1}$~cm$^{-2}$ (0.3-10~keV) or $F_{\nu,X}=0.13\,\mu$Jy at 1.7~keV.

We calculate the value of $p$ from both indices for two scenarios: $\nu_m < \nu_X < \nu_c$, where $\nu_m$ is the peak frequency of the synchrotron spectrum, and $\nu_X > \nu_c$; requiring that the value of $p$ needs to be in agreement for each scenario. We find that $\nu_X > \nu_c$, with a weighted mean and 1$\sigma$ uncertainty of $\langle p \rangle$ = 2.01 $\pm$ 0.15. Since typical values of $p$ range between 2-3 due to implications that arise from the distribution of the Lorentz factors (e.g. \citealt{Jager1992}), we employ $p$ = 2.1 in subsequent analysis.

Since the X-rays lie above $\nu_c$, the isotropic-equivalent kinetic energy does not depend on the circumburst density, and thus can be used to constrain $E_{\rm{K,iso}}$ directly, assuming fixed values for $\epsilon_{e}$ and $\epsilon_{B}$. Fixing $z$ = 1.754, $D_L$ = 13457 Mpc, $\nu_X$ = 1.7 keV (the logarithmic centre of the 0.3-10 keV XRT band), $\epsilon_e=\epsilon_B=0.1$ (c.f., \citealt{Zhang2015}), we use $F_{\nu,X}=0.13\,\mu$Jy at 0.10~days to calculate
\begin{equation}
    E_{\rm{K,iso},52} = 0.14 \pm 0.03,
\end{equation}
where $E_{\rm{K,iso},52}$ is $E_{\rm{K,iso}}$ in units of $10^{52}$ erg.
An additional constraint can be set in the limiting case that $\nu_c$ is at the lower edge of the X-ray band, $\nu_{\rm c,max}$ = 0.3 keV, which places a lower limit on the combination of $E_{\rm{K,iso}}$ and $n$, of
\begin{equation}
    n^{2}E_{\rm{K,iso},52} > 5.99 \times\ 10^{-5}
\end{equation}
where $n$ is in units of cm$^{-3}$.
For the optical and NIR bands, we assume that $\nu_m < \nu_{\rm{opt/NIR}} < \nu_c$. From $i=25.10 \pm 0.39$ at 0.38~days, we calculate $F_{\nu,{\rm opt}}=0.33 \pm 0.14 \,\mu$Jy, and obtain
\begin{equation}
    n^{0.4} E_{\rm{K,iso},52} = 0.03 \pm 0.01.
\end{equation}
and the NIR constraint of $F_{\nu,{\rm NIR}} < 1.87 \,\mu$Jy at 0.43 days gives us
\begin{equation}
    n^{0.4} E_{\rm{K,iso},52} < 0.32.
\end{equation}
Finally, we use the available 9~GHz ATCA upper limit at 0.52 days \citep{gcn23467}, and make the assumption $\nu_{sa} < \nu_{\rm{radio}} < \nu_m$, where $\nu_{sa}$ is the self-absorption frequency, to calculate
\begin{equation}
    n^{0.6} E_{\rm{K,iso},52} < 0.2.
\end{equation}
For the case where $\nu_{\rm{radio}} < \nu_{sa}$, we find no difference in the cumulative allowed parameter space, which is primarily determined by the optical and X-ray detections and the cooling frequency constraint.

The allowed $E_{\rm{K,iso}}$-$n$ parameter space for GRB\,181123B, calculated from combining the probability distributions from the above relations, is shown in Figure \ref{fig:energy_density}. We calculate the medians for the parameters, $E_{\rm{K,iso,52}} = 0.13^{+0.02}_{-0.02}$ erg and $n = 0.04^{+0.02}_{-0.01}$ cm$^{-3}$, from the 1D probability distributions\footnote{For these parameters, we find a global Compton $Y\approx0.1$ \citep{se01}, suggesting that inverse-Compton (IC) cooling is not significant. Whereas IC cooling becomes more important for lower values of $\epsilon_B$, in practice, the inclusion of Klein-Nishina corrections severely limits the maximum Compton Y-parameter for faint bursts, especially at low density \citep{Nakar2009}. Thus, we do not include IC cooling in our analysis.} We also calculate the above constraints for $\epsilon_B = 0.01$, finding $E_{\rm{K,iso,52}} = 0.14^{+0.02}_{-0.02}$ erg and $n = 1.10^{+0.87}_{-0.32}$ cm$^{-3}$ (Figure~\ref{fig:energy_density}). Motivated by the low value of $\epsilon_B\approx 10^{-4}-10^{-2}$ derived for GW170817's afterglow (e.g., \citealt{Hajela2019}, \citealt{Wu2018}), as well as those derived for {\it Swift} SGRBs \citep{Santana2014,Zhang2015}, we explore the possibility of a low value of $\epsilon_B$, and find that the allowed parameter space is completely ruled out for $\epsilon_B \lesssim 10^{-3}$ (for $\epsilon_e=0.1$). In summary, we derive $E_{\rm K, iso} \approx 0.13-0.14 \times\ 10^{52}$~erg and $n \approx 0.04-1.10$~cm$^{-3}$ for GRB\,181123B. Using the value of $E_{\rm K,iso}$ and $E_{\gamma{\rm ,iso}}=5.0 \times 10^{51}$~erg\footnote{We note that because we do not model the gamma-ray spectrum in this work, we do not know the true bolometric correction, but have used a fiducial value of 5.}, we also calculate a gamma-ray efficiency of 0.78, just above the median of 0.57 found by \cite{Fong2015} and in line with the the higher values found by \citet{Beniamini2016b} when no Synchrotron Self-Compton component is included.

\section{Host galaxy properties} \label{sec:host}
To characterize the host galaxy of GRB\,181123B, we use \texttt{Prospector} (\citealt{Leja2017}, \citealt{Johnson2017}), a stellar population modeling code which employs a library of Flexible Stellar Population Synthesis models (FSPS; \citealt{Conroy2009}, \citealt{Conroy2010}) and determines the best-fit solution and posterior parameter distributions with \texttt{Dynesty} \citep{Speagle2020} through a nested sampling algorithm \citep{Skilling2004,Skilling2006}. We fit our photometric data to independently determine the redshift, $z_{\rm{photo}}$, as well as the following stellar population properties: rest-frame attenuation in mags ($A_V$), stellar metallicity (Z), mass ($M_{\star}$ in units of solar mass), star formation history (SFH), and age of the galaxy at the time of observation, $t_{\rm age}$. During fits, these parameters can either be set free to determine the posterior distribution or fixed to a specific value and adopt priors that are uniform across the allowed parameter space within FSPS. For the SFH, we employ a parametric delayed-$\tau$ model, such that SFR(t) $\propto te^{-t/\tau}$, with $\tau$ as an additional free parameter. We then use $t_{\rm age}$ and $\tau$ to convert to a mass-weighted age of the galaxy, $t_{\rm{gal}}$, by $t_{\rm{gal}} = t_{\rm{age}} - \frac{\int_{0}^{t_{\rm{age}}} \rm{SFR}(t) t dt}{\int_{0}^{t_{\rm{age}}} \rm{SFR}(t) dt}$. We use a Chabrier initial mass function (IMF, \citealt{Chabrier2003}), a Milky Way extinction law (R = $3.1/\rm{E}(B-V)$, \citealt{Cardelli1989}), turn nebular emission on to model a SF galaxy, and add additional attenuation towards the nebular regions to account for the fact that stars in SF regions will generally experience twice the attenuation of normal stars \citep{Calzetti2000, Price2014}. 

\begin{figure}
\includegraphics[width=0.475\textwidth]{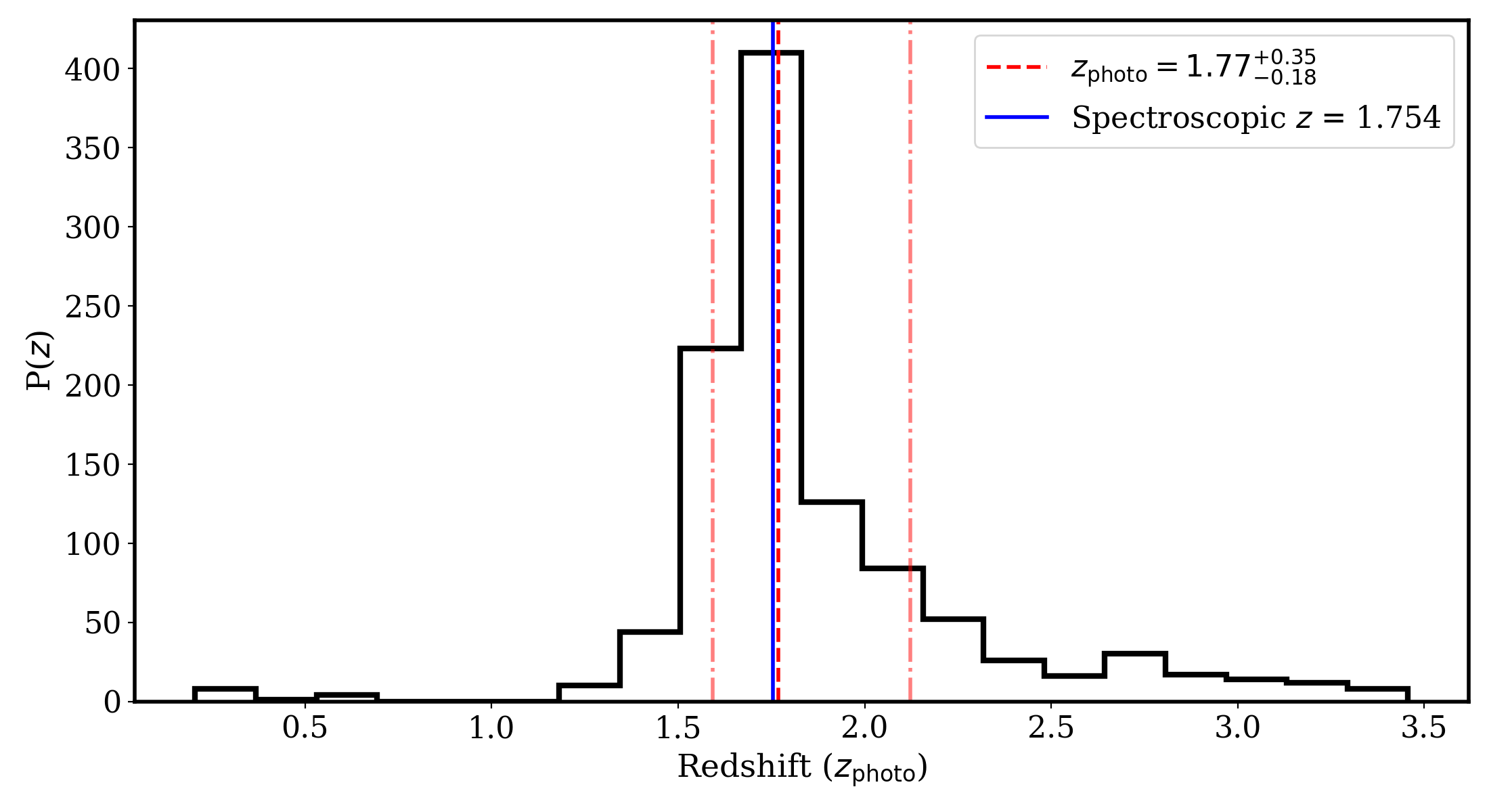}
\caption{Posterior distribution of the redshift, $z_{\rm{photo}}$, found by \texttt{Prospector} using the photometric data of GRB\,181123B's host galaxy, over the final 1085 iterations. We find a single peak of $z_{\rm{photo}}$ = $1.77^{+0.30}_{-0.17}$, fully consistent with the spectroscopically determined redshift, assuming the single emission line is H$\beta$. \label{fig:prospector_z}}
\end{figure}

First, we perform a fit to determine the photometric redshift, $z_{\rm{photo}}$, using the $grizYJHK$ photometry and 1$\sigma$ uncertainties of GRB\,181123B's host galaxy, along with the relevant transmission curves for each filter (obtained from the corresponding website of each instrument\footnote{\url{https://www2.keck.hawaii.edu/inst/mosfire/throughput.html}}\footnote{\url{https://www.cfa.harvard.edu/mmti/mmirs/instrstats.html}}\footnote{\url{https://www.gemini.edu/instrumentation/gmos/components\#Filters}}, \citealt{Crampton2000,McLean2012,McLeod2012}). We allow a range of $z = 0.2-4$, and Z, $A_V$, $\tau$, $t_{\rm{age}}$, and $M_{\star}$ to be additional free parameters. We find a single-peaked posterior distribution for the redshift using the final 1085 iterations of the sampling once the solution has converged (Figure \ref{fig:prospector_z}), with $z_{\rm{photo}}$ = $1.77^{+0.30}_{-0.17}$. This is consistent with the redshift determined from the single emission line identified in the NIR spectrum if the line is H$\beta$. There is a low-probability tail ($<$ 0.1) extending to high redshifts, but solutions beyond $z \approx 2.5$ are inconsistent with the spectrum and photometric colors.

\begin{figure}
\includegraphics[width=0.49\textwidth]{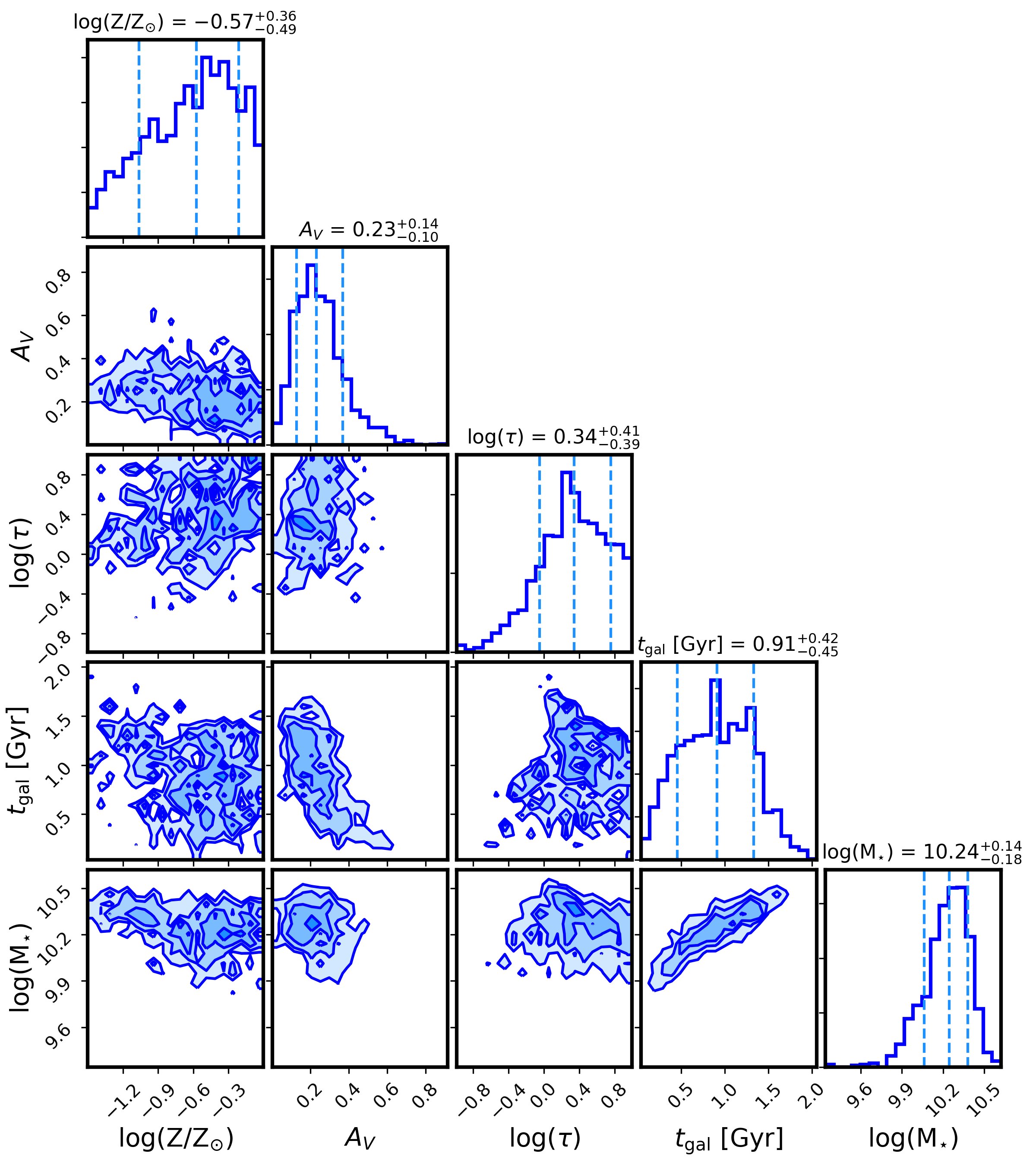}
\caption{Corner plots showing the fitted parameters found by \texttt{Prospector} using the photometric data of GRB\,181123B's host galaxy using the spectroscopic $z$ = 1.754. \label{fig:prospector_fit}}
\end{figure}

\begin{figure*}
\centering
\includegraphics[width=0.8\textwidth]{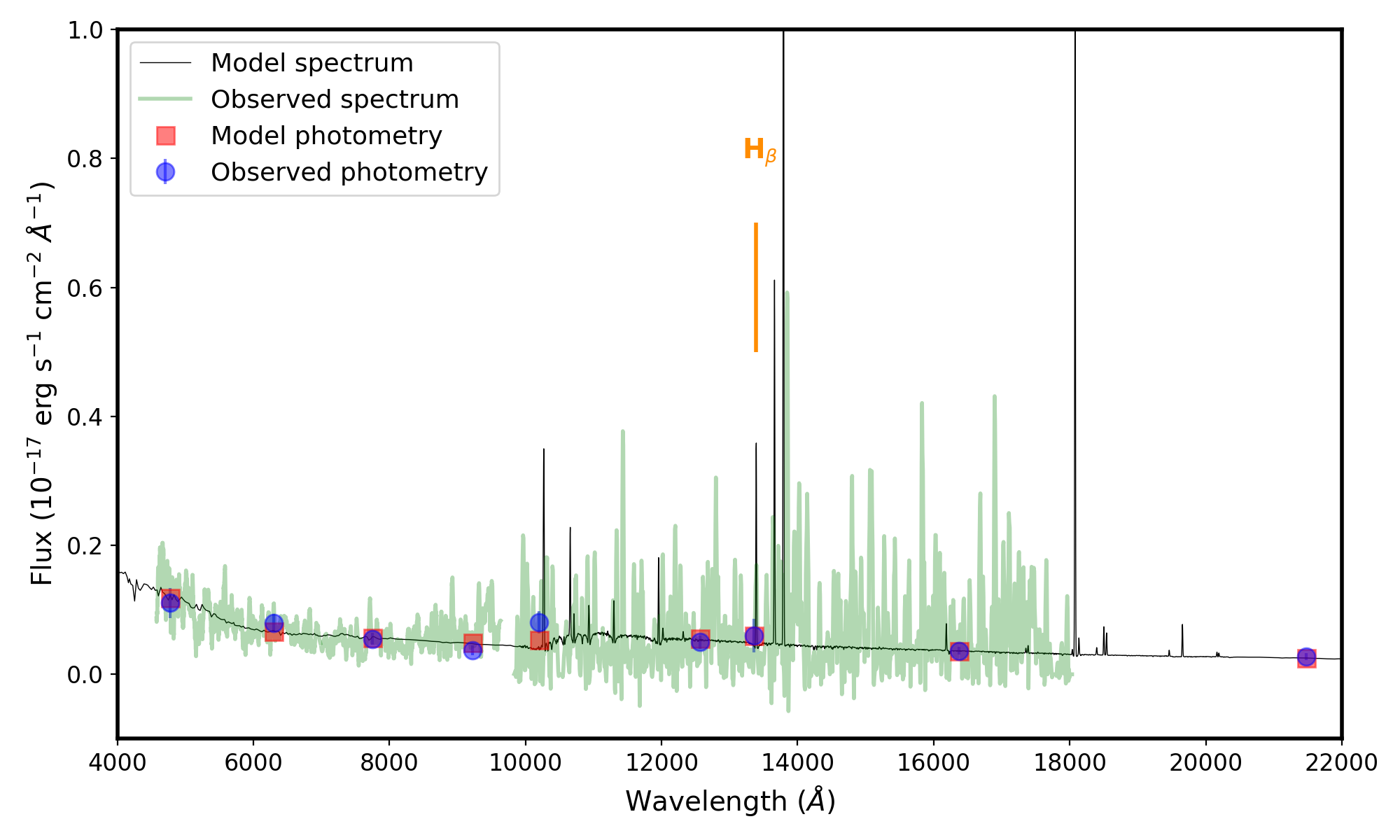}
\caption{The photometry and spectrum of GRB\,181123B's host as a function of the observed wavelength, overplotted on the model spectrum and model photometry calculated from \texttt{Prospector} at $z$ = 1.754. The orange line shows the position of the single emission line identified in the NIR spectrum. The model spectrum and photometry provide good agreement with the shape of the spectral continuum and the photometric colors. \label{fig:prospector_SED}}
\end{figure*}

Next, we fix the redshift to the spectroscopically-determined value of $z = 1.754$, set the maximum value of $t_{\rm age}$ to be the age of the Universe at that redshift (3.755 Gyr) and fit for the remaining stellar population properties. To self-consistently account for attenuation while calculating the SFR, we also include an additional synthetic photometric data point calculated from the spectrum by defining a box filter of 300 \AA~ width centred on the H$\beta$ emission line. We find final values of log(Z/$Z_{\odot}$) = $-0.57^{+0.36}_{-0.49}$, $A_V$ = $0.23^{+0.4}_{-0.10}$ mag; log($\tau$) = $0.34^{+0.41}_{-0.39}$ Gyr; and log($M_{\star}$) = $10.24^{+0.14}_{-0.16}$ M$_{\odot}$. Calculating the mass-weighted age from the star formation history gives $t_{\rm gal} = 0.91^{+0.42}_{-0.45}$ Gyr. The corner plot produced by \texttt{Prospector}, showing the parameter posterior distributions, is shown in Figure \ref{fig:prospector_fit}, while the observed photometry (including the synthetic photometric point around H$\beta$), overplotted with the model spectrum and photometry, as well as the observed spectrum is shown in Figure \ref{fig:prospector_SED}. 
We note that there are well-known degeneracies between $A_V$, $Z$ and $t_{\rm age}$ \citep{Conroy2013}. We explore these degeneracies by fixing metallicity and rest-frame attenuation to a range of values to see how they affect the parameter solutions. Except when $A_V$ is set to the extreme cases of 0 or 1 mag, which produces an unconstrained $t_{\rm{age}}$ and a poor fit to the data respectively, the remaining parameters solutions remain within a narrow range of values.

Scaling by the total mass formed, we find a SFR = $32.82^{+16.34}_{-7.24}$ M$_{\odot}$ yr$^{-1}$ and log(sSFR) = $-8.72^{+0.26}_{-0.19}$ yr$^{-1}$ from the SED. In principle, we can also use the NIR emission line to determine a SFR by calculating a H$\alpha$ flux using relations from \cite{Kennicutt1998}. This method however, is subject to stellar absorption and dust attenuation and relies on typical H$\beta$/H$\alpha$ line ratios, which can lead to deviation from the true SFR by several factors \citep{Moustakas2006}. Indeed, without correction, we determine a SFR$_{\rm{H}\beta}$ = $4.91 \pm 0.43$ M$_{\odot}$ yr$^{-1}$, $\sim$ 6 times lower the SED SFR. We therefore consider the SFR calculated from the SED to be true representation of the SFR.

\section{Discussion} \label{sec:discussion}
\subsection{Comparing GRB 181123B to the SGRB Population}
At $z=1.754$, GRB\,181123B is among the most distant SGRBs with a confirmed redshift to date. Comparing the $\gamma$-ray $T_{90}$, hardness ratio and fluence to those of {\it Swift} SGRBs across $z=0.1-2.2$ \citep{Lien2016}, GRB\,181123B lies near the median value compared to the rest of the population, solidifying its membership in this class. GRB\,111117A originates from a host galaxy at a higher redshift of $z=2.211$, and also has similar $\gamma$-ray properties to those of other SGRBs \citep{Selsing2018}. In contrast, the high-redshift GRB\,090426A at $z=2.609$ has a measured $T_{90} \sim 1.3$~sec which would ostensibly place it in the SGRB class, but has spectral and energy properties which are otherwise more similar to those of long GRBs \citep{Antonelli2009,Levesque2010}, so the classification and progenitor of this burst is unclear and we do not include it in our discussion of secure SGRBs. GRB\,051210 is likely at $z>1.4$ due to its featureless optical host galaxy spectrum \citep{Berger2007}, but does not have a secure redshift. Finally, GRB\,160410A has an afterglow redshift of $z=1.717$ \citep{Selsing2019} and is most likely a SGRB with extended emission \citep{gcn19276}. This makes GRB\,181123B one of a few SGRBs with a confirmed redshift at $z>1.5$, and the highest-redshift secure SGRB to date with an optical afterglow detection.

We next examine the inferred afterglow and host galaxy properties of GRB\,181123B in the context of the SGRB population. In Table~\ref{tab:comparison}, we present several properties for GRB\,181123B as well as where the event falls in the SGRB population as a percentile; in this scheme, a value of $50\%$ is the median value of that parameter. At the most basic level, the faint apparent magnitude of the optical afterglow ($i \approx 25.1$ at $\sim 9.1$~hours) puts GRB\,181123B in the lower 30\%. However, when corrected for the redshift of the burst, GRB\,181123B's afterglow luminosity is slightly above the median of other SGRBs at similar rest-frame times.

The detection of both the X-ray and optical afterglows of GRB\,181123B allows us to constrain the isotropic-equivalent kinetic energy scale and circumburst density to $E_{\rm{K,iso}} \approx 1.3 \times 10^{51}$~erg and $n \approx 0.04-1.1$~cm$^{-3}$. For a direct comparison to GRB\,111117A, we determine the allowed $E_{\rm K,iso}-n$ parameter space in the same manner as described in Section~\ref{sec:burst} using the X-ray afterglow detection and optical upper limit \citep{Margutti2012,Sakamoto2013} at $z=2.211$, finding $E_{\rm K,iso}=(1.4-2.3) \times 10^{51}$~erg and $n=0.0045-0.13$~cm$^{-3}$ where the range is set by $\epsilon_B=0.01-0.1$. While the kinetic energy scales for both bursts are similar to those of SGRBs, with a median value of $\approx 2\times 10^{51}$~erg \citep{Fong2015}, the inferred circumburst density of GRB\,181123B is at the higher end of the population (Table~\ref{tab:comparison}). Recently, \citet{Wiggins2018} used cosmological simulations and population synthesis models for BNS mergers to predict the circum-merger densities as a function of redshift. Overall, they found that the fraction of mergers occurring in high-density environments increases with redshift, with the median density changing from $\approx 10^{-3}$~cm$^{-3}$ at $z<0.5$ to $\approx 0.1$~cm$^{-3}$ at $z>1$. While the larger circumburst density of GRB\,181123B seems to align with this predicted trend, we note that the other bursts with inferred densities of $>0.1$~cm$^{-3}$ predominantly originate at low redshifts of $z<0.5$. Moreover, the expectation is for high circumburst densities to correspond to smaller offsets (modulo projection effects), but the projected physical offset of GRB\,181123B is $5.08 \pm 1.38$~kpc, just below the population median of $\approx 6$ kpc (Fong et al., in prep). While the number of high-redshift bursts is admittedly too small at present for robust comparisons, based on the current sample at $z>1.5$, we do not find any appreciable trends between SGRB afterglow properties and redshift.

\begin{figure}[t]
\includegraphics[width=0.47\textwidth]{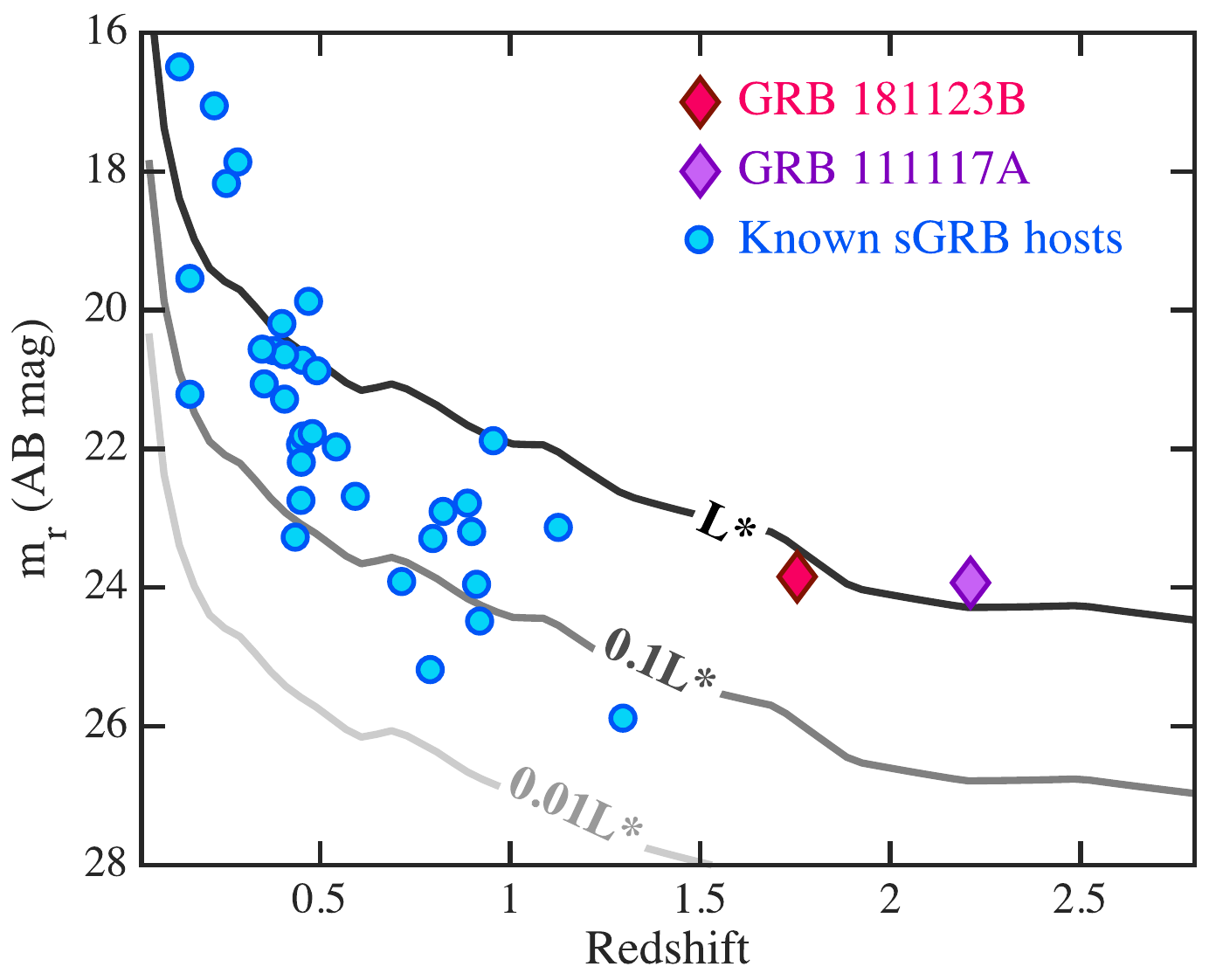}
\caption{Host galaxy apparent $r$-band magnitude ($m_r$) of 34 SGRBs with known redshifts and optical measurements (blue circles). The high-redshift events GRBs\,181123B (this work) and 111117A \citep{Selsing2018} are highlighted as diamonds. Contours denote the evolving galaxy luminosity function across redshift corresponding to $L^*$, $0.1L^*$ and $0.01L^*$. Both GRBs\,181123B and 111117A are $\sim L^*$ galaxies compared to those at contemporary redshifts (see text). \label{fig:lumz}}
\end{figure}

To understand how GRB\,181123B fits in the context of SGRB hosts, we collect data for 34 SGRBs with known redshifts and measured apparent $r$-band magnitudes of their host galaxies ($m_r$) from the literature \citep{LeiblerBerger2010,Levesque2010,Fong2013,Troja2016,Fong2017,Selsing2018,Lamb2019,Selsing2019} and our own observations. We compare the values of $m_r$ to the characteristic luminosity, $L^*$ across redshift, using available galaxy luminosity functions \citep{Brown2001,Wolf2003,Willmer2006,ReddySteidel2009,Finkelstein2015}. For each redshift, we take the value of $L^*$ in the band that corresponds to the observed $r$-band, blue-shifted to its rest-frame at that redshift. We then interpolate across redshift to create smooth contours corresponding to $L^*$, $0.1L^*$, and $0.01L^*$. A comparison of the SGRB host population to the evolving galaxy luminosity function is shown in Figure~\ref{fig:lumz}. At $z \lesssim 1$, SGRB hosts span the luminosity range of $\approx 0.05-1L^*$ while at $z \gtrsim 1.5$, they are on the upper end of the luminosity function. This trend can be easily explained by observational bias as only the more luminous galaxies will be detectable at higher redshifts. At $L \approx 0.9L^{*}$, the host galaxy of GRB\,181123B is similar to that of GRB\,111117A ($1.2L^{*}$; Figure~\ref{fig:lumz} and \citealt{Selsing2018}). In terms of stellar mass ($10^{10.24}\,M_{\odot}$) and (mass-weighted) stellar population age (0.9~Gyr), the host properties of GRB\,181123B are also typical of the SGRB population, which has median values of $\approx 10^{10.24}\,M_{\odot}$ and 1.07~Gyr (Nugent et al., in prep; Table~\ref{tab:comparison}). The redshift and stellar population age of GRB\,181123B implies that $50\%$ of its stellar mass was formed when the universe was $\approx 2.8$~Gyr old, corresponding to $z \sim 2.3$, around the peak of the cosmic SFR density \citep{MadauDickinson2014}. 

Next, we compare the host galaxy of GRB\,181123B to the expected properties for galaxies at $z \approx 1.5-2$. The rest-frame $U-V$ and $V-J$ colors have long been used to distinguish quiescent from SF galaxies to $z\sim2$ \citep{Williams2009}. At $z=1.754$, rest-frame $U-V$ is roughly equivalent to $z-H$ or $Y-H$, which we calculate to be $\approx1.2$ and $\approx1.1$~mag, respectively for GRB\,181123B's host galaxy. This places the host galaxy in the region occupied by unobscured SF galaxies in the $UVJ$ diagram, for all possible $V-J$, at $1.5<z<2.0$ \citep{Fumagalli2014}. Using the values of SFR $\approx$ 33 $M_{\odot}$~yr$^{-1}$ and log(sSFR) $\approx$ $-8.7~$yr$^{-1}$ derived from the SED, we find that the host of GRB\,181123B lies on the SF main sequence (SFMS) for galaxies of similar mass at the same redshift \citep{Whitaker2014,Fang2018}; and find a similar result for the host of GRB\,111117A based on the $H\alpha$-derived SFR \citep{Selsing2018}.

\begin{deluxetable}{cccc}
\tablecaption{Comparison of properties of GRB\,181123B \label{tab:comparison} }
\tablecolumns{3}
\tablewidth{0pt}
\tablehead{
\colhead{Properties} &
\colhead{GRB\,181123B} &
\colhead{SGRBs$^{\diamond}$} &
\colhead{GRB\,111117A} 
}
\startdata
$z$ & 1.754 $\pm$ 0.001 & 98\% & 2.211 \\
$T_{90}$ (s) & 0.26 $\pm$ 0.04 & 33\% & 0.46 $\pm$ 0.05\\
Hardness & 2.4 $\pm$ 0.6 & 78\% & 2.8 $\pm$ 0.5\\
$E_{\gamma{\rm , iso,52}}$ (erg) & 0.50 & 79\% & 0.86\\
$E_{\rm{K,iso,52}}^{\dagger}$ (erg) & 0.13-0.14 & 39\% & $0.14-0.23$ \\
$n^{\dagger}$ (cm$^{-3}$) & 0.04-1.10 & 72-95\% & $0.005-0.13$\\
$L$ ($L^{*}$) & 0.9 & $65\%$ & 1.2\\
log($M_\star$) (M$_{\odot}$) & 10.24$^{+0.14}_{-0.18}$ & 50\% & 9.9\\
Age (Gyr) & 0.91$^{+0.42}_{-0.45}$ & 44\% & 0.5$^\ddagger$\\
Proj. offset (kpc) & $5.08 \pm 1.38$ & 44\% & $10.52 \pm 1.68$\\
\enddata
\tablecomments{$^{\diamond}$ Percentile for GRB\,181123B compared to SGRB population \\
$^{\dagger}$ Values assuming $\epsilon_{B}$ = 0.01-0.1 \\
$^{\ddagger}$ This is the derived SSP age, so can be taken as a lower limit on the true age of the stellar population (c.f., \citealt{Conroy2013}). \\
Values for GRB\,111117A are taken from \citet{Selsing2018} and \citet{Lien2016}, except for $E_{\rm K,iso}$ and $n$ which are derived in this work. SGRB comparison samples are from \citet{Fong2015}, \citet{Fong2017}, and Nugent et al. (in prep.).}
\end{deluxetable}

\subsection{SGRB Redshift Distribution and Implications for Delay Times}

\begin{figure*}[t]
\centering
\includegraphics[width=0.95\textwidth]{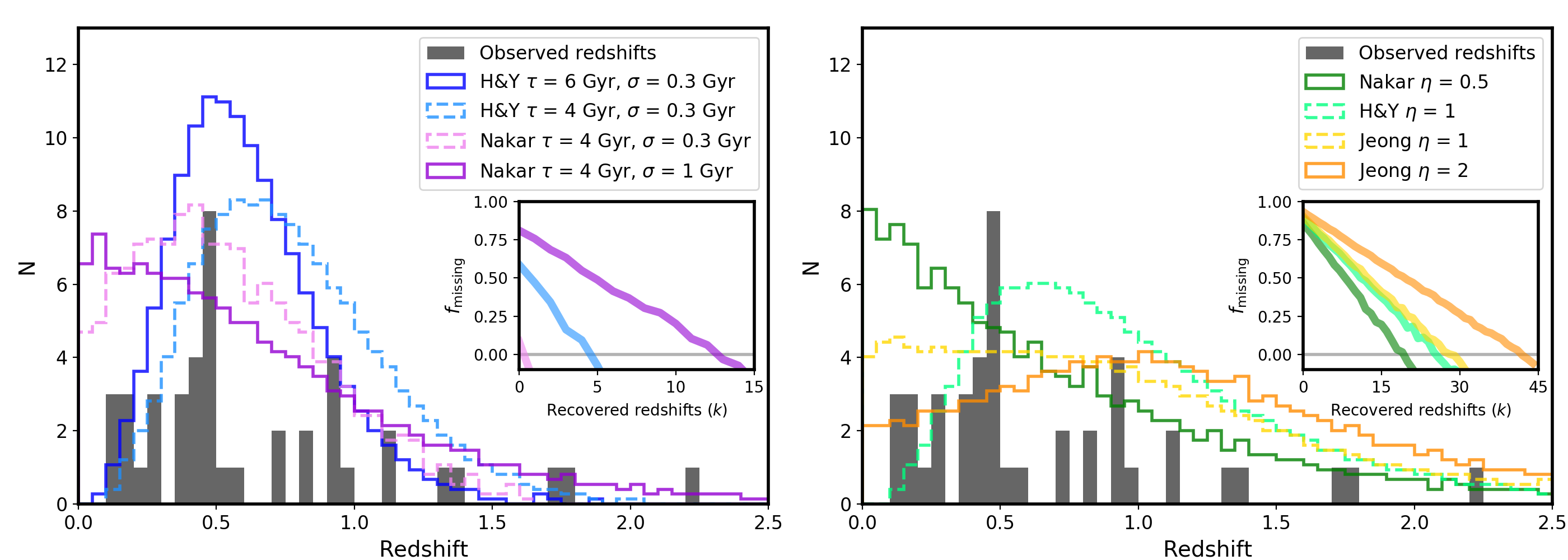}
\caption{Redshift distribution of the 43 SGRBs with known redshifts (solid black line), plotted with the predicted observed redshift distributions derived from log-normal (left, \citealt{Nakar2006,Hao2013}) and power-law (right, \citealt{Nakar2006,Jeong2010,Hao2013}) DTDs, representing the entire {\it Swift} SGRB population (134 events). The log-normal models favor lower redshifts, while the power-law models allow for more high-$z$ bursts. The inset in each plot shows the $95\%$ confidence upper limit on the missing fraction of high-$z$ bursts ($z > 1.5$) as a function of the number of high-$z$ bursts recovered ($k$) for each model. The grey horizontal line at a missing fraction of zero denotes the point when the respective model is ruled out to 95\% confidence. As expected, the log-normal models which favor low redshifts are quickly ruled out with a few bursts recovered at high-$z$, while the power-law models can accommodate a larger number of high-$z$ bursts.} \label{fig:redshift}
\end{figure*}

With the detection of GRB\,181123B, there are only $3$ {\it Swift} SGRBs with confirmed redshifts at $z>1.5$ (and only $7$~SGRBs at $z>1$). The apparent lack of SGRBs at high redshifts can be attributed to a combination of (i) {\it Swift} detector insensitivity at high redshifts (e.g., \citealt{GuettaPiran2005,Behroozi2014}), (ii) the difficulty of obtaining secure redshifts in the so-called `redshift desert' ($1.4 < z < 2.5$) in which all strong nebular emission lines are redshifted to $>1~\mu$m and Ly$\alpha$ is not yet accessible, and (iii) the intrinsic delay time distribution (DTD), imprinted from a neutron star merger progenitor \citep{Belczynski2006}. In this section, we explore the constraints we can place from the observed SGRB redshift distribution on DTD models, with a focus on high-$z$ ($z>1.5$) events.

In the context of their binary merger progenitors, the true fraction of SGRBs which originate at $z>1.5$ has implications for the merger timescale distribution, binary progenitor properties (e.g., initial separations, eccentricities; \citealt{Selsing2019}), and $r$-process element enrichment which in turn can have effects on galaxy properties across redshift \citep{OShaughnessy2010,OShaughnessy2017,Safarzadeh2019,Simonetti2019}. Of particular importance are the delay times, the time interval encompassing the stellar evolutionary and merger timescales, which impact our understanding of compact binary formation channels (isolated binary evolution versus dynamical assembly in globular clusters). The two main functional forms that have been widely considered in the literature are a power-law DTD (characterized by $t^{-\eta}$) and a log-normal DTD (characterized by mean delay time $\tau$ and width $\sigma$). The observed SGRB distribution peaks at $z\approx 0.5$ with a steep drop-off toward higher redshifts \citep{Berger2014,Fong2017}, ostensibly favoring a log-normal DTD with long delay times of several Gyr \citep{Hao2013}. This functional form could be explained by dynamical formation in globular clusters, in which the delay time strongly depends on the relaxation timescale of the cluster for NS binaries to assemble, which can be several Gyr \citep{Spitzer1987,Hopman2006,Kremer2019,Ye2019}. On the other hand, a power-law DTD naturally arises for primordial binaries (i.e. systems that were born as a pair and have co-evolved) given a power-law distribution of initial orbital separations and coalescence due to gravitational wave losses \citep{Peters1964,OShaughnessy2007,Dominik2012}.  Indeed, Type Ia SN studies have found that the observations are consistent with a DTD described by a power-law with $\eta=1$ \citep{Totani2008,Maoz2012,Graur2014}.
 
Thus far, studies of the Galactic binary neutron star population \citep{VignaGomez2018} as well as SGRB host galaxy demographics are in rough agreement with power-law DTDs with $\eta \gtrsim 1$ \citep{Zheng2007,FongBerger2013}. We note that some studies have found an excess of more rapid mergers, finding steeper DTDs than $\eta=1$ \citep{BeniaminiPiran2019}, but each provide fairly weak constraints. If we are indeed missing a population of high-$z$ SGRBs, this would indicate overall shorter delay times and provide an additional constraint on the DTD. Some studies have also suggested a possible bimodal DTD distribution \citep{Salvaterra2008}, but this is in tension with more recent theoretical studies showing that dynamical assembly of NS-NS and NS-BH mergers can only contribute a small fraction to the overall merger rates \citep{Belczynski2018,Ye2020}.

The current observed SGRB redshift distribution comprises 43 events (updated from \citealt{Fong2017}), out of a total 134 {\it Swift} SGRBs detected to date \citep{Lien2016}. This sample comprises all SGRBs with a secure host association ($P_{cc}<0.1$), and a spectroscopic afterglow or host redshift, or a well-sampled photometric host redshift. This serves as an initial basis for comparison to predicted redshift distributions with varying underlying DTDs, star formation histories and luminosity functions. Much work has been done in the literature to perform the convolution between these functions and predict the observed redshift distributions \citep{GuettaPiran2005,Nakar2006,Hao2013,Wanderman2015,Anand2018}. From these works, we collect eight representative predicted distributions from the literature that cover the entire observed SGRB redshift range ($z \sim 0.1-2.5$), and are not already ruled out by current observations. Four models describe log-normal DTDs with $\tau = 4$ and $6$~Gyr, and widths of $\sigma=0.3$ and $1$~Gyr \citep{Nakar2006,Hao2013}. The remaining four models are power-law DTDs with $\eta = 0.5-2$ \citep{Nakar2006,Jeong2010,Hao2013}. We note that models with the same DTD parameters may give rise to slightly different distributions due to the underlying assumptions on the star formation histories, luminosity functions, which can result in significant changes, and detector sensitivity (Figure~\ref{fig:redshift}).

First, we use two-sample Kolmogorov-Smirnov (K-S) statistics to test the null hypothesis that each model is consistent with being drawn from the same underlying distribution as the observed redshift distribution of 43 SGRBs. With the exception of the log-normal distribution with $\tau=4$~Gyr ($\sigma=0.3$~Gyr) from \citet{Hao2013} which predicts a peak in the distribution at $z \approx 0.75$, all of the log-normal distributions are consistent with being drawn from the same underying distribution as the observed data, and we cannot reject the null hypothesis ($p=0.45-0.62$). On the other hand, all of the power-law distributions with $\eta\geq 1$ result in $p<0.05$, and thus are not consistent with being drawn from the same underlying distribution, while the power-law DTD characterized by $\eta=0.5$ is consistent ($p=0.3$). It is clear that without taking into account observational biases, distributions dominated by long delay times are preferred \citep{Hao2013}.

This analysis, however, neglects the inherent biases in the observed SGRB redshift distribution. Thus, we explore the fraction of {\it Swift} SGRBs that could be originating at high-$z$ among the {\it current} population. Previous studies have found that $\approx 33-70\%$ of SGRBs could be missing at redshifts of $z>0.7-1$ \citep{Berger2007,Selsing2018}. Over 2004-2020, {\it Swift} has detected 134 SGRBs (including 13 with extended emission; \citealt{Lien2016}), and 43 have secure redshift determinations ($P_{\rm cc}<0.1$; 31\%). In the large majority of cases, the determination of a redshift depends on an association to a host galaxy which requires precise positional information from the detection of an X-ray or optical afterglow ($\lesssim$~few arcsec precision). In the case of an afterglow detection, the lack of redshift can be attributed to the lack of a coincident host galaxy due to significant kicks and merger timescales, leaving large displacements between the burst and host galaxy \citep{Berger2010,Fong2013b,Tunnicliffe2014}; offsets of $\gtrsim 10$~kpc are predicted to comprise as much as $40-50\%$ of the total population \citep{Wiggins2018}\footnote{We note that large kicks are at odds with those inferred from the Galactic BNS population; cf. \citealt{BeniaminiPiran2016,Tauris2017}}. A faint host galaxy may also escape detection due to a low-luminosity or high-$z$ origin \citep{OConnor2020}. In this case, an apparently-faint galaxy is more likely to originate at lower redshifts due to the increase in the faint-end slope of the galaxy luminosity function \citep{Blanton2005,Parsa2016}, although Figure~\ref{fig:lumz} shows that SGRB hosts overall are drawn from the brighter end of the galaxy luminosity function. In total, the number of SGRBs which lack redshift information is 91 events. If we assume that $50\%$ of these events arise at $z>1$, this translates to $\approx 34\%$ ($46$ events) of the current {\it Swift} SGRB population. If we take into account the 34 bursts that were subject to constraints which prevent the detection of an afterglow and subsequent redshift determination, such as satellite observing constraints, poor sight-lines, or high Galactic extinction and follow the same arguments as above, this results in $\approx 28\%$ (28 bursts) of the current population that were not subject to significant observing constraints. These numbers can be directly compared to expectations from DTDs.

We perform an exercise to explore how many SGRBs need to be recovered at high-redshifts before a given model can be ruled out (``recovered redshifts''). We concentrate here on high-redshifts given that these have comparatively larger discriminating power between DTD models than low-redshift events. To convert each of the eight continuous model distributions into a representative redshift distribution, we draw 1000 events from each model and then scale to 134 events (Figure~\ref{fig:redshift}). For each model, we determine the fraction of SGRBs which could originate at $z>1.5$ as predicted by the model. To account for counting statistics, we draw 134 bursts from each model distribution 1000 times, and determine the 95\% confidence region on the high-$z$ fraction. We then compute the missing fraction demanded by each model as a function of the number of additional, recovered redshifts at $1.5 < z< 3$ ($k$), taking into account that there are already 3 known SGRBs at $z>1.5$ so the observed population would be $3+k$. In Figure~\ref{fig:redshift}, when the missing fraction goes to $0$, the model can be ruled out to $95\%$ confidence.

We find that log-normal models with small widths of $\sigma=0.3$~Gyr can be ruled out for as few as $k\lesssim 1-5$ recovered redshifts, while the wider width, $\sigma=1$~Gyr model, could still accommodate $k=13$~additional events recovered at $z>1.5$. However the shape of the low-redshift distribution is not supported by this model (Figure~\ref{fig:redshift}). For the power-law distributions, we find a significantly larger number of high-$z$ bursts are allowed before the models are ruled out to $95\%$ confidence ($k \approx 19-42$ recovered redshifts), although models with $\eta=0.5$ and $\eta=2$ significantly under- or over-predict the $z<1$ population. We find similar results if we perform the same analysis with a population of 100 events (representing the {\it Swift} SGRB population with no observing constraints). Performing a K-S test on each of the new distributions assuming 100 events shows agreement with the 134 event results. With the addition of high-$z$ bursts to the observations, the data quickly favor the power-law distribution with $\eta = 1$ and all log-normal distributions are ruled out by the null hypothesis. Our analysis shows that the SGRB population is more consistent with power-law DTD models with $\eta=1$, and the recovery of only a few high-$z$ bursts, together with the shape of the low-$z$ population, will help to solidify the model parameters. From the $\eta=1$ power-law model, we find that the expected number of SGRBs which originate at $z>1$ is $\approx 45$ events ($33\%$ of the current population). Compared to our estimate that $\approx 34\%$ of {\it Swift} SGRBs originate at $z>1$, this is another line of support for the $\eta=1$ power law model, and thus a primordial NS binary formation channel.

In Type Ia SNe studies, similar work has been done to constrain the `prompt' fraction. Indeed, \citet{Rodney2014} found that observations suggest a prompt fraction of up to 50\% (defined as events with delay times of $<500$~Myr), with the results fully consistent with a power-law DTD with $\eta$ = 1. For NS mergers, most recent simulations require a prompt channel to explain $r$-process enrichment in Milky Way stars and ultra-faint dwarf galaxies \citep{Matteucci2014,Beniamini2016,Simonetti2019,Safarzadeh2019d}. A reliable estimate of the prompt SGRB fraction would require a careful assessment of observational biases, the true SGRB redshift distribution, and stellar population ages as a proxy for the progenitor age distribution. Nevertheless, additional, future detections of SGRBs at $z\approx 2$ and beyond might help to quantify the true prompt fraction of SGRBs.

\section{Conclusions} \label{sec:conclusion}
We have presented the discovery of the optical afterglow and host galaxy of GRB\,181123B at $z=1.754$, contributing to a small but growing population of SGRBs at high redshifts. These results are based on a rapid-response and late-time follow-up campaign with Gemini, Keck, and the MMT. Our main conclusions are as follows:
\begin{itemize}
    \item GRB\,181123B is the second most distant bona-fide SGRB with a confirmed redshift measurement, after GRB\,111117A ($z=2.211$). It is the most distant SGRB to date with an optical afterglow detection.
    \item The host galaxy of GRB\,181123B is characterized by a stellar mass $\approx 1.7 \times 10^{10}\,M_{\odot}$, luminosity of $\approx 0.9L^*$ and mass-weighted age of $\approx 0.9$~Gyr. These are comparable to the median values of the SGRB host population across redshift.
    \item Compared to the SF main sequence of galaxies at $1.5<z<2.0$, GRB\,181123B lies just below or significantly below this sequence, and is thus forming stars at a lower rate than most SF galaxies of similar mass, indicating that it is moving toward quiescence.
    \item The current redshift distribution comprises 43 events, and is consistent with most log-normal distributions with moderate delay times ($\approx 4-6$~Gyr). An analysis of the full {\it Swift} sample of 134 events, taking into account the difficulty of confirming high-$z$ SGRBs, demonstrates that log-normal DTD models are overall disfavored. In particular, models with moderate delay times of $\approx 4-6$~Gyr and small widths of $\sigma=0.3$~Gyr can be ruled out to $95\%$ confidence with an additional $\lesssim 1-5$~{\it Swift} SGRBs recovered at $z>1.5$. Log-normal models with wider widths of $\sigma = 1$~Gyr are less favored given the lack of low-$z$ SGRBs.
    \item Power-law DTDs with an index around unity are more consistent with the data and can accommodate $\approx 30$ recovered SGRBs at $z>1.5$ ($22\%$ of the current population). For this model $\approx 45$ of the remaining SGRBs are expected to have $z>1$ ($33\%$ of the current population.) This is consistent with our estimates on the observed fraction of SGRBs originating at $z>1$ of $\approx 34\%$, and is also consistent with SGRBs originating from primordial NS binaries.
\end{itemize}

In order to properly constrain the DTD and probe the underlying formation channels of SGRBs and BNS mergers, it is important to uncover high-$z$ bursts ($z>1.5$). However, high-$z$ bursts provide additional challenges for follow-up due to the additional observations needed and the resources available.  The determination of the redshift of GRB\,181123B required $6$ to $10$-meter class telescopes, and highlights the sheer difficulty of obtaining redshifts for host galaxies at $z>1.5$, where the main spectral features are redshifted to near-IR wavelengths with no major features at optical wavelengths. Moreover, even if SGRBs are drawn from the brighter end of the galaxy luminosity function, the host magnitudes are still $J\approx 22-23$~mag. Dedicated efforts to characterize high-$z$ candidates among the current population with state-of-the-art NIR instruments, as well as the planned {\it JWST} and ELTs will help to solidify the true high-$z$ fraction among the current population. In the era of gravitational wave multi-messenger astronomy, NS mergers detected via gravitational waves may also help constrain the DTD through the studies of their host galaxies and connecting the redshift distributions of BNS mergers to those of SGRBs (e.g., \citealt{Safarzadeh2019a,Safarzadeh2019b,Safarzadeh2019c}).

\section*{Acknowledgements}
We acknowledge Sarah Wellons, Allison Strom, David Sand for helpful discussions that aided this work, and to Mansi Kasliwal for facilitating Keck observations. The Fong Group at Northwestern acknowledges support by the National Science Foundation under grant Nos. AST-1814782 and AST-1909358. This work was also in part supported by the National Aeronautics and Space Administration through grant HST-GO-15606.001-A from the Space Telescope Science Institute, which is operated by the Association of Universities for Research in Astronomy, Incorporated, under NASA contract NAS5-26555, and Chandra Award Number DD7-18095X issued by the Chandra X-ray Center, which is operated by the Smithsonian Astrophysical Observatory for and on behalf of NASA under contract NAS8-03060. MN is supported by a Royal Astronomical Society Research Fellowship. A.J.L has received funding from the European Research Council (ERC) under the European Union's Horizon 2020 research and innovation programme (grant agreement No. 725246, TEDE, PI Levan). A.A.M.~ is funded by the Large Synoptic Survey Telescope Corporation, the
Brinson Foundation, and the Moore Foundation in support of the LSSTC Data Science Fellowship Program; he also receives support as a CIERA Fellow by the CIERA Postdoctoral Fellowship Program (Center for Interdisciplinary Exploration and Research in Astrophysics, Northwestern University). J.L. is supported by an NSF Astronomy and Astrophysics Postdoctoral Fellowship under award AST-1701487. K.D.A. acknowledges support provided by NASA through the NASA Hubble Fellowship grant HST-HF2-51403.001-A awarded by the Space Telescope Science Institute, which is operated by the Association of Universities for Research in Astronomy, Inc., for NASA, under contract NAS5-26555.

Based on observations obtained at the international Gemini Observatory (PIs Paterson, Fong; Program IDs GS-2018B-Q-112, GN-2018B-Q-117, GS-2019A-FT-107), a program of NOIRLab, which is managed by the Association of Universities for Research in Astronomy (AURA) under a cooperative agreement with the National Science Foundation on behalf of the Gemini Observatory partnership: the National Science Foundation (United States), National Research Council (Canada), Agencia Nacional de Investigaci\'{o}n y Desarrollo (Chile), Ministerio de Ciencia, Tecnolog\'{i}a e Innovaci\'{o}n (Argentina), Minist\'{e}rio da Ci\^{e}ncia, Tecnologia, Inova\c{c}\~{o}es e Comunica\c{c}\~{o}es (Brazil), and Korea Astronomy and Space Science Institute (Republic of Korea).

W. M. Keck Observatory and MMT Observatory access was supported by Northwestern University and the Center for Interdisciplinary Exploration and Research in Astrophysics (CIERA). Some of the data presented herein were obtained at the W. M. Keck Observatory (PIs Miller, Fong, Paterson; Programs 2018B\_NW254, 2018B\_NW249, 2019A\_O329), which is operated as a scientific partnership among the California Institute of Technology, the University of California and the National Aeronautics and Space Administration. The Observatory was made possible by the generous financial support of the W. M. Keck Foundation. The authors wish to recognize and acknowledge the very significant cultural role and reverence that the summit of Maunakea has always had within the indigenous Hawaiian community.  We are most fortunate to have the opportunity to conduct observations from this mountain. Observations reported here were obtained at the MMT Observatory, a joint facility of the University of Arizona and the Smithsonian Institution (PI Fong, Programs 2018C-UAO-G4, 2019A-UAO-G7, 2020A-UAO-G212-20A).

This research was supported in part through the computational resources and staff contributions provided for the Quest high performance computing facility at Northwestern University which is jointly supported by the Office of the Provost, the Office for Research, and Northwestern University Information Technology.

This work made use of data supplied by the UK \textit{Swift} Science Data Centre at the University of Leicester.

\vspace{5mm}
\facilities{\textit{Swift}(XRT and UVOT), Gemini-North (GMOS), Gemini-South (GMOS, FLAMINGOS-2), Keck:I (MOSFIRE), Keck:II (DEIMOS), MMT (MMIRS)}

\bibliographystyle{aasjournal}
\bibliography{biblio}

\end{document}